**RESEARCH ARTICLE**

# 3D Relativistic MHD simulations of the gamma-ray binaries

M.V. Barkov,[1] E. Kalinin,[1,2] and M. Lyutikov[3]

[1]Institute of Astronomy, Russian Academy of Sciences, Moscow, 119017 Russia
[2]Moscow Institute of Physics and Technology, Dolgoprudny, 141701 Russia
[3]Department of Physics and Astronomy, Purdue University, West Lafayette, IN 47907-2036, USA
**Author for correspondence:** M.V. Barkov, Email: barkov@inasan.ru.

**Abstract**

In gamma-ray binaries neutron star is orbiting a companion that produces a strong stellar wind. We demonstrate that observed properties of "stellar wind"-"pulsar wind" interaction depend both on the overall wind thrust ratio, as well as more subtle geometrical factors: the relative direction of the pulsar's spin, the plane of the orbit, the direction of motion, and the instantaneous line of sight. Using fully 3D relativistic magnetohydrodynamical simulations we find that the resulting intrinsic morphologies can be significantly orbital phase-dependent: a given system may change from tailward-open to tailward-closed shapes. As a result, the region of unshocked pulsar wind can change by an order of magnitude over a quarter of the orbit. We calculate radiation maps and synthetic light curves for synchrotron (X-ray) and Inverse-Compton emission (GeV-TeV), taking into account $\gamma - \gamma$ absorption. Our modeled light curves are in agreement with the phase-dependent observed light curves of LS5039.

## 1. Introduction

Gamma-ray binaries are an important class of high energy astrophysical sources (Guillaume Dubus 2013). A canonical example is LS 5039, which has historically been the subject of intense multiwavelength campaigns (Hadasch et al. 2012; Collmar and Zhang 2014) (other sources include (LS I +61 ˚ 303, HESS J0632+057, 1FGL J1018.6-5856 and 1FGL J1018.6-5856).

In these sources, the relativistic pulsar wind interacts with the wind of the companion and later with the interstellar matter. In the process, ultra-relativistic particles are accelerated and emit radio emission to gamma rays (see, e.g., Tavani and Arons 1997; Sierpowska and Bednarek 2005; G. Dubus 2006; Khangulyan et al. 2007; Kong, Cheng, and Huang 2012; Zabalza et al. 2013; Dubus, Lamberts, and Fromang 2015; Molina and Bosch-Ramon 2020; Huber, Kissmann, and Reimer 2021; Khangulyan, Barkov, and Popov 2022; Lopez-Miralles et al. 2022). The accretion black-hole as a compact object was also discussed in the literature(see J. Casares et al. 2005; Bosch-Ramon and Khangulyan 2009; Barkov and Khangulyan 2012), but it looks less feasible due to the strict energetic constraint observed in the MeV energy range (Collmar and Zhang 2014).

Importantly, the interaction of the wind from the neutron star with the wind from a high-mass companion leads to the *orbital-dependent* X-ray emission and gamma-ray emission. Often flares are observed at particular orbital phases (*e.g.* Abdo et al. 2011). Understanding the origin and properties of this orbital dependence is the main goal of the present work.

There is general agreement that this complex behavior is the result of the wind-wind interaction between the relativistic highly magnetized wind of neutron star and a powerful (and anisotropic) wind from the main sequence companion (Guillaume Dubus 2013), the main mechanisms of the non-thermal emission are synchrotron and Inverse Compton (see Bosch-Ramon and Khangulyan 2009), but the details of the interaction and production of γ-ray emission remain controversial.

Here we point out that the wind-wind interaction depends both on the global properties of the pulsar (its spin-down power) and of the companion's wind (mass and momentum loss rates), as well on the more subtle geometrical properties: relative direction of the pulsar's rotational axis and its magnetic inclination, and orbital-dependent direction from the pulsar to the companion. In addition, since effects of relativistic beaming are likely to be important, the view of the interacting region depends on the orbital-dependent line of sight. This requires both 3D simulations, as well as investigation of various parameters.

The system resembles the case of the pulsar-ISM interaction studied by Barkov, Lyutikov, and Khangulyan 2019, but is also different in many ways. In the case of pulsar-ISM interaction we can identify two basic geometric types depending on the relative orientation of the pulsar rotational axis and the velocity: (i) a "Rifle Bullet", with the pulsar spin and velocity aligned; (ii) a "Frisbee", with the pulsar spin and velocity orthogonal to each other. The internal dynamics of the shocked pulsar wind is considerably different in these cases. The "Frisbee" configuration results in substantially *non-symmetric morphologies of the tail region* due to the influence of the internal hoop stresses.

Somewhat similar configurations are expected in the binary case, Fig. 1. One difference of the wind-wind interaction from the wind-ISM case is that the relative geometry is orbital phase dependent (except for "Frisbee" configuration). For the case of spin in the plane of the orbit, the system evolves from the "Bullet" to the "Cartwheel" configuration every quarter of the period.

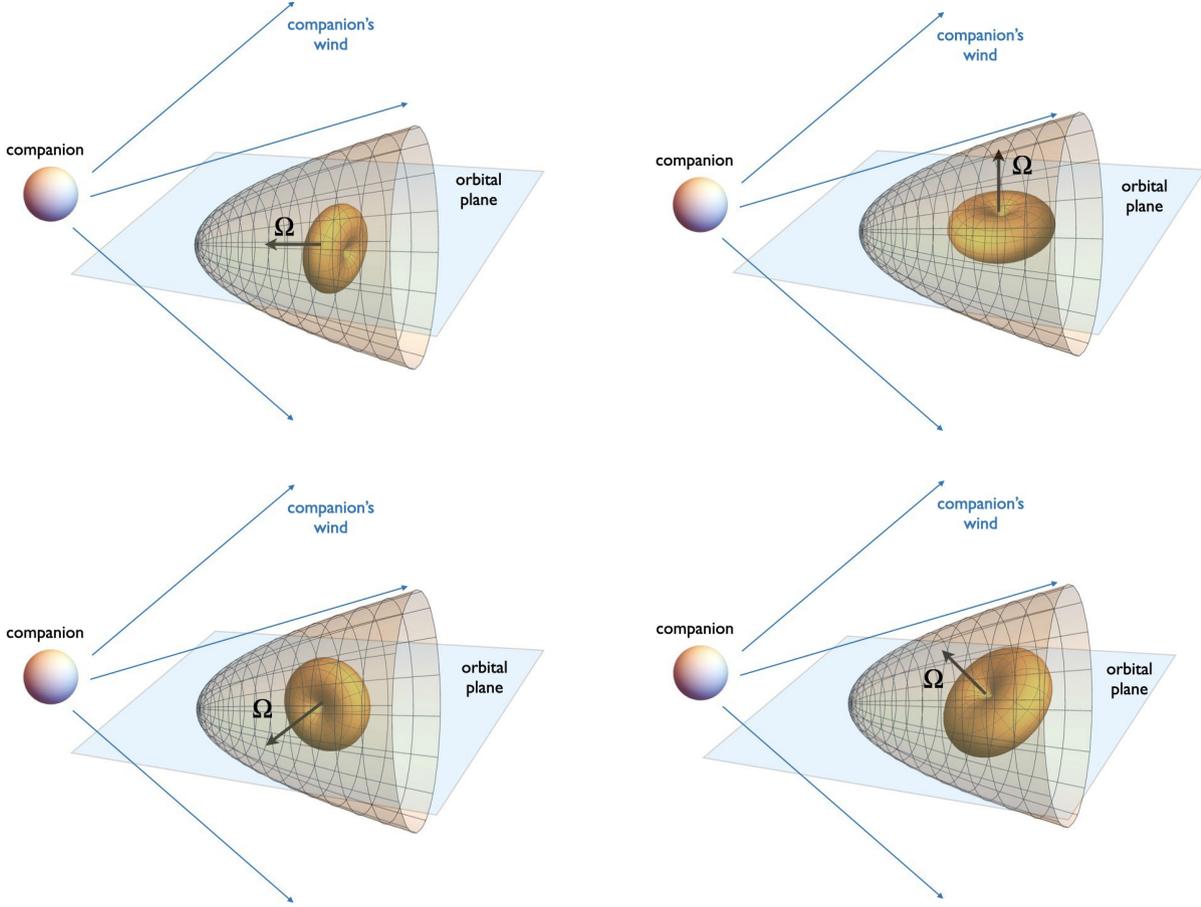

**Figure 1.** Basic geometries: "Rifle Bullet" (the spin of the neutron star is in the orbital plane, directed towards the companion, top left panel), "Frisbee" (with the spin of the neutron star perpendicular both to the orbital plane, top right panel), "Cartwheel" (spin of the neutron star in the orbital plane but perpendicular to the direction to the companion, bottom left panel), and a mixed "Frisbee-Bullet" configuration (bottom right panel). The central doughnut-like structure indicates the distribution of wind power.

The most important difference between the wind-ISM and wind-wind interaction is the tail structure of the resulting pulsar wind nebulae (PWN). First, the winds' collision creates two shocks separated by contact discontinuity (CD). The shock in the normal star wind we call the forward shock (FS) and the second one in the pulsar wind we call the reverse shock (RS), Fig. 2. In the wind-ISM case, the tailward pulsar wind always terminates at the so-called Mach disk – a strong, nearly perpendicular shock (perpendicular in a sense that flow velocity is nearly orthogonal to the shock plane Barkov, Lyutikov, and Khangulyan 2019). In the wind-wind case "open" structure has a possibility of creating – when the weaker wind extends as a supersonic flow to infinity in the tail-ward direction, see Fig. 2 left panel and Fig. 10 of Bogovalov et al. 2008. In such a case, the confined pulsar wind expands conically, keeping the force balance with the confining wind.

Conventionally, in a fluid description (Bogovalov et al. 2008), the key parameter of the interaction of winds is momentum thrust ratio $\eta$ which in the case of pulsar and stellar wind interaction is

$$\eta = \frac{L_{sd}}{\dot{M}v_w c}, \quad (1)$$

here $\dot{M}$ and $v_w$ are stellar wind mass loss rate and speed correspondingly, $c$ is speed of light and $L_{sd}$ is pulsar spindown power. For sufficiently small $\eta \leq 0.01$ the Mach disc appears at large distances down the pulsar tail (see Fig. 2 right panel), of the order of the orbital separation, and moves to shorter distances if we reduce $\eta$. Thus, we expect a sensitive dependence of the overall morphology, and consequently of the emission properties, on the momentum thrust ratio $\eta$.

As we demonstrate in this paper, the open-closed dichotomy also depends on the orbital-dependent geometrical properties of the pulsar wind, *e.g.* Fig. 3. The same system may evolve into different configurations depending on the orbital phase. On the Fig. 3 we indicated a position of the reverse shock in pulsar wind by orange line.

In order to understand the complicated orbital-dependent behavior of these systems, we performed a set of relativistic 3D MHD simulations of the interaction of the relativistic pul-sar wind with the stellar wind. We rely on (and extend) two previous related investigations: hydrodynamic models of the wind-wind interaction in gamma-ray binaries (Bogovalov et al. 2008; Bogovalov et al. 2012; V. Bosch-Ramon et al. 2012; Bosch-Ramon, Barkov, and Perucho 2015; Valenti Bosch-Ramon et al. 2017; Barkov and Bosch-Ramon 2016, 2021; Lamberts et al. 2013; Dubus, Lamberts, and Fromang 2015; Huber, Kissmann, and Reimer 2021) and 3D relativistic MHD simulations of the pulsar wind-ISM interaction (Barkov, Lyutikov, and Khangulyan 2019; Barkov et al. 2019; Olmi and Bucciantini 2019).

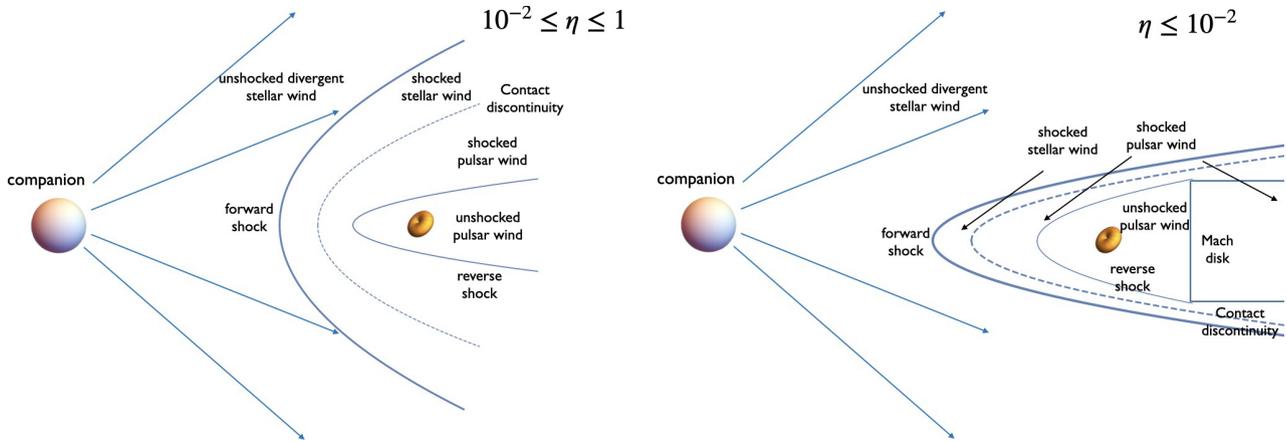

**Figure 2.** Two possible structures of the tail flow: "open" one, when the supersonic pulsar wind partially extends to infinity, and "closed" one, when the supersonic pulsar wind always terminates at the reverse shock. For a given thrust ratio η the type of configuration also depends on the neutron star geometry, see Fig. 3. In the case pulsar-ISM interaction only closed configurations are possible. For "Stellar wind"-"pulsar wind" case switching between the two morphologies leads to sudden changes in the resulting emission.

In the case of fluid simulations, V. Bosch-Ramon et al. 2012; Bosch-Ramon, Barkov, and Perucho 2015 found that the wind-wind interaction is subject to strongly non-linear processes already within the orbital scale that leads to the isotropization of the interaction region, and loss of coherence. Eventually the interacting region becomes an irregular isotropic flow formed by mixed stellar and pulsar winds. This mixed flow terminates with a shock on the external medium (interstellar medium (ISM)) or the interior of a supernova remnant (SNR).

Near the the bow-shock region, one can adopt the interaction of a magnetized pulsar wind with the interstellar medium (ISM) as an approximate model for the interaction of a pulsar and a stellar wind. Recently, two papers (Barkov, Lyutikov, and Khangulyan 2019; Barkov et al. 2019) have presented a study of the formation of pulsar wind nebulae from a fast-moving pulsar. These works show that the geometry head of the bow-shock and the pulsar-wind tail can be significantly affected by the formation of the jet-like structures along pulsar spin axis. The contact discontinuity in the pulsar wind tail can take the form of a cross (in "Frisbee" case), instead of a cylindrical shape, as in the case of a spherically symmetric nonmagnetized wind moving fast and interacting with the ISM. This cross shape of the contact discontinuity is the result of the asymmetry of the pulsar wind, which is dominated by the equatorial flow and the jet-like structures in the polar directions.

The paper has the following structure: 1) Introduction; 2) Details of the simulation setup (stellar wind and pulsar wind setup); 3) Simulation results with a detailed comparison with 2D RHD and 2D RMHD[a] and a detailed discussion of the flow in general; 4) Discussion and conclusions.

## 2. Details of the simulations' setup.

The initial setup of our simulation for the magnetized pulsar wind is similar to our previous work for the interaction of the ISM with the relativistic wind of a fast moving pulsar (Barkov, Lyutikov, and Khangulyan 2019, see also Porth, Komissarov, and Keppens 2014). As a first step, we focus on the bow-shock structure and study the formation of tail-ward Mach disks, neglecting the orbital motion of the pulsar.

The simulations were performed using a three-dimensional (3D) geometry in Cartesian coordinates using the *PLUTO* code[b] (Mignone et al. 2007). Spatial linear interpolation, a 2nd order Runge-Kutta approximation in time, and an HLL Riemann solver were used (Harten 1983). *PLUTO* is a modular Godunov-type code entirely written in C and intended mainly for astrophysical applications and high Mach number flows in multiple spatial dimensions. The simulations were performed on CFCA XC50 cluster of National Astronomical Observatory of Japan (NAOJ). The flow has been approximated as an ideal, relativistic adiabatic gas, one particle species, and polytropic index of $4/3$. The size of the domain is $x \in [-2, 10]$, $y$ and $z \in [-5, 5]$. We have uniform resolution in the entire computational domain with total number of cells $N_X = 780$, and $N_Y = N_Z = 650$, see details in Table 1.

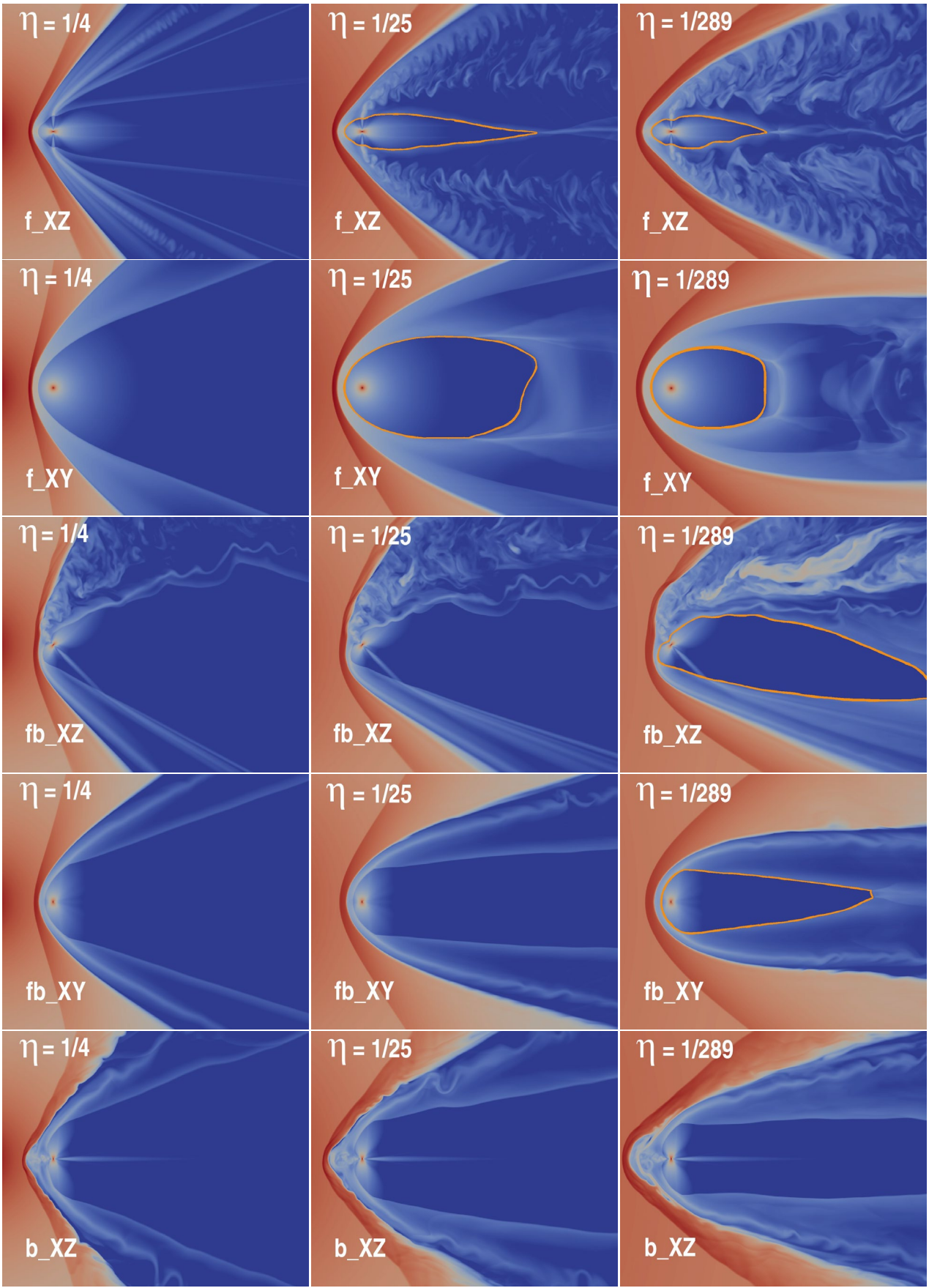

**Figure 3.** Density (color) for the models with $\eta = 1/4$ (left column), $\eta = 1/25$ (central column), and $\eta = 1/289$ (right column). At the first row "Frisbee" XZ-plane and the second row XY-plane, the "Frisbee-Bullet" XZ-plane and XY-plane on the third and fourth rows respectively, and the "Bullet" XZ-plane on the fifth row. The position of the reverse shock in pulsar wind is indicated by orange line.

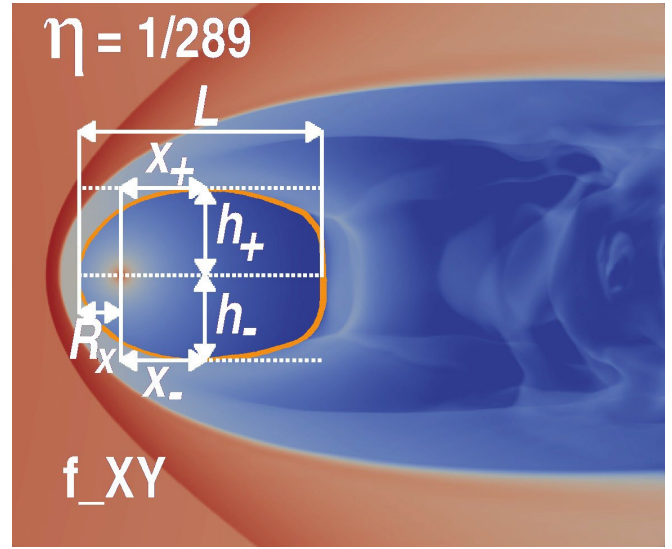

**Figure 5.** Sketch with close wind zone, which may be in situations with "Frisbee" and "Frisbee-Bullet" orientations. The measured parameters for the models are presented in the Tab.4.

### 2.1 Stellar wind setup

The stellar wind is simulated as an already formed, full-speed, supersonic radial flow centered in the star location carrying toroidal magnetic field. To make the simulation computationally lighter, we use a non-uniform resolution in the computational domain. All these assumptions allow us to use a much larger resolution at the region encompassing the pulsar and its bow shock, as well as a large linear size of the computational domain.

In our model the normal star has place outside the computational domain, because of supersonic inflow on the left X edge ($X = -2$) we can inject such the stellar wind. For models r3 we set the normal star position at the point (-3,0,0), for r6 at the point (-6,0,0), and r18 at the point (-18,0,0).

We start our simulation from a non-equilibrium configuration and evolve it up to the time at which quasi-stationary solution is reached. To reduce computational expenses, we set the stellar wind speed as $v_{\rm wind} = 0.1c$ and Mach number $M = 85$. The density of the stellar wind was adopted so that in the case of non-magnetized spherical pulsar wind, the bow-shock is formed at the position near (–1, 0, 0). The wind speed value is not realistic, but it is not significantly affecting the volume inside the contact discontinuity (Barkov, Lyutikov, and Khangulyan 2020). The stellar wind have weak magnetization $\sigma_{\rm wind} = 0.01$, which is formed by toroidal/ axisymmetric (Y-axis) magnetic field in the plane-XZ.

This setup has several free parameters which we vary in our simulations: (i) changing the pulsar distance to the normal star while keeping the distance to the bow shock constant (in dimensional less units, actual distance from pulsar to standing point in the reverse shock is function of $\eta$) allows us to inspect how sensitivity is the numerical solution to variations of

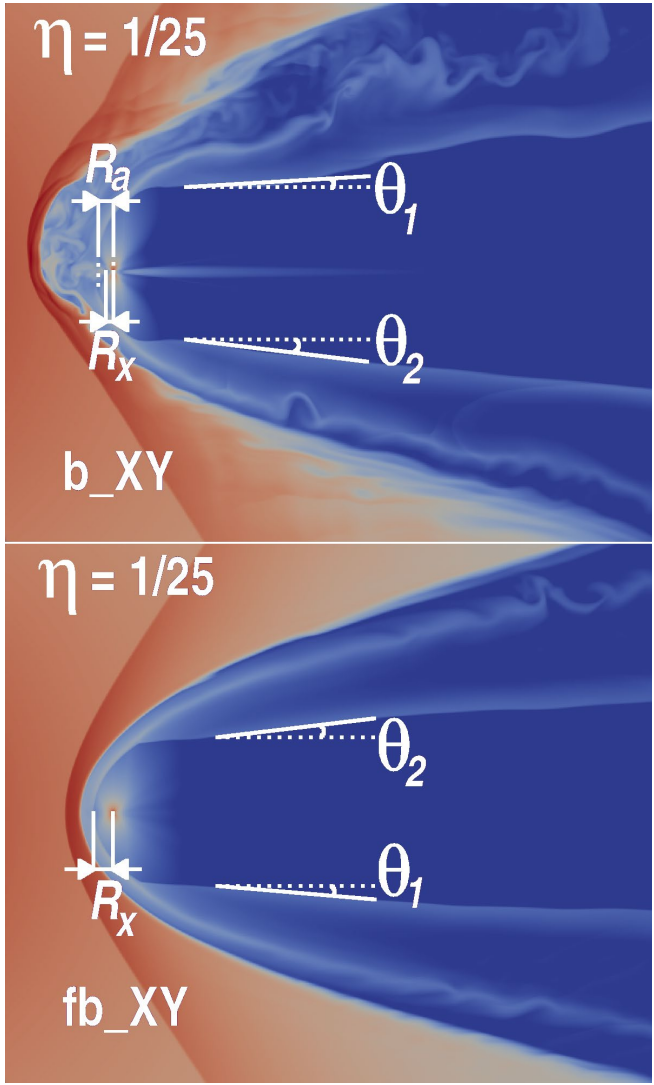

**Figure 4.** Sketches of two different configurations with open free wind zone. The upper case may be only in "Bullet" orientation, and the lower case may be in "Frisbee" or "Frisbee-Bullet" orientation. The measured parameters for the models are presented in the Tab.3.

**Table 1.** Parameters of the computational Grid

| Coordinates | Left | $N$ | Right |
|---|---|---|---|
| $X$ | –2 | 780 | 10 |
| $Y$ | –5 | 650 | 5 |
| $Z$ | –5 | 650 | 5 |

---

a. In the hydrodynamical cases without orbital motion 2D and 3D simulations are very similar.
b. Link http://plutocode.ph.unito.it/index.html

the non-dimensional momentum-rate ratio, the pulsar wind magnetization, and the orientation of pulsar rotation axis (i.e. the magnetic field toroidal geometry) relative to the direction to the normal star; (ii) we study different spin orientations of the pulsar with respect to the wind velocity (as the pulsar moves along the orbit the orientation of the pulsar spin change with respect to the wind direction, except in the case when pulsar spin axis is perpendicular to the orbital plane); (iii) we include the rotation of the pulsar spin axis along the orbit, which mimics the effect of orbital motion.

### 2.2 Pulsar wind setup

The pulsar emits the unshocked magnetized pulsar wind with a toroidal magnetic field, which changes its polarity in the northern and southern hemispheres.

In our work we use the prescription of pulsar wind described in the paper (Porth, Komissarov, and Keppens 2014), see details in the Appendix 1. The pulsar with radius 0.15 is placed at the point (0,0,0). The pulsar wind was injected with the initial Lorentz factor $\Gamma = 1.7$ and Mach number 25. For all models, we adopt the same magnetization parameter $\sigma_0 = 1$[c] and angle between the magnetic and rotation axis $\alpha = \pi/4$. The thrust ratio of the pulsar and stellar wind was chosen to form a bow shock at the distance of 1 for the spherical pulsar wind. Parameters for the models are presented in the Table 2.

We choose three cases of pulsar orientation "Frisbee", "Bullet" and intermediate one "Frisbee-Bullet" (Barkov, Lyutikov, and Khangulyan 2019). In the case of the "Frisbee" the pulsar rotation axis is parallel to the axis Z ($\zeta = 0$), in the case of the "Bullet" the pulsar rotation axis is parallel to the axis X ($\zeta = \pi/2$), the intermediate case was formed by clockwise turn of the "Frisbee" configuration around the Y axis at angle $\zeta = \pi/4$. Such change of configurations is natural if the pulsar rotation axis lays in the orbital plane.

We performed calculations for nine models, as we vary the η parameter and pulsar orientation in respect to direction to the normal component, Table 2. These static cases can be interpreted as evolution of the spin axis in the binary system if the pulsar spin lies in the orbital plane. Thus, neglecting the influence of the magnetic field in the stellar wind (which was chosen small and dynamically not important) we can also interpret our results as different individual binary systems with various orientation of spin axis in respect to orbital plane. It can be systems with three sequences: 1) if the pulsar spin axis is normal to the orbital plane, we have "Frisbee" all the time; 2) if the pulsar spin axis angled on 45° to the orbital plane, we have transition "Frisbee" — "Frisbee-Bullet" — "Frisbee" every half of the orbit; and 3) If the pulsar spin axis lay in the orbital plane, we have "Cartwheel" — "Frisbee-Bullet" — "Bullet" — "Frisbee-Bullet" — "Cartwheel" every half of the orbit.

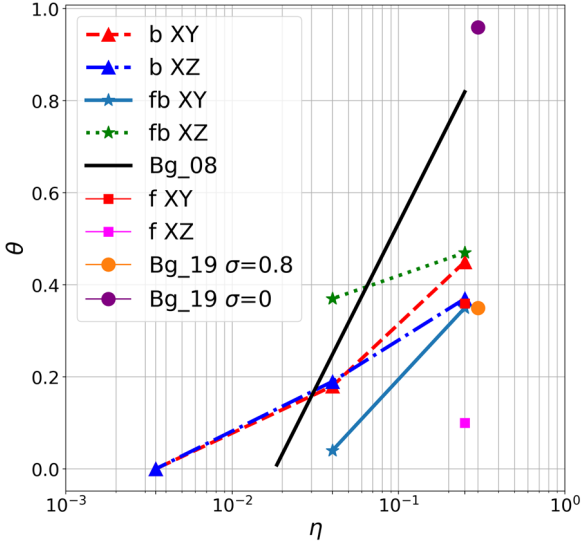
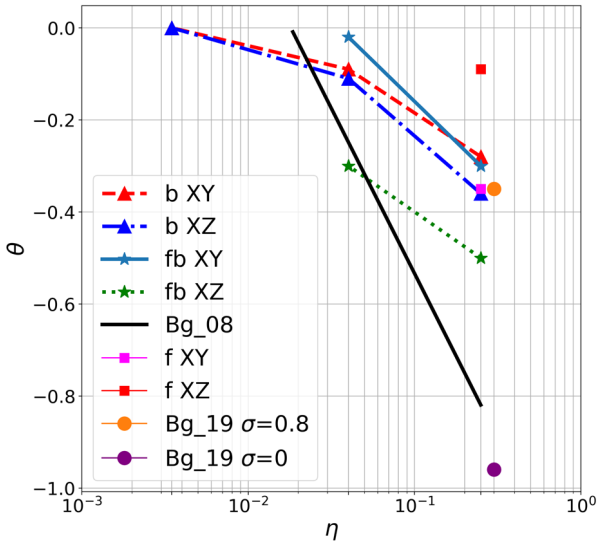

**Figure 6.** Dependence of angles $\theta_1$ (upper) and $\theta_2$ (lower) on the eta parameter $\eta$. The "Frisbee", "Bullet" and "Frisbee-Bullet" orientations were marked by 'squares', 'triangles' and 'stars' correspondingly. The measured parameters for the models are presented in the Tab.3. The black thick solid line denotes the result obtained in the work (Bogovalov et al. 2008). The orange and purple dots denote the opening angles for magnetization σ 0.8 and 0 correspondingly obtained in the work (Bogovalov et al. 2019).

### 3. Results

---

c. In the paper of Bogovalov et al. 2019 was used $\sigma_0 = 0.8$, but the difference is not critical (see Fig.6).

**Table 2.** Parameters of the models. 1) Name of the model; 2) angle between direction to the star and pulsar spin axis; 3) stellar and pulsar wind thrust ratio; 4) distance to the normal star in the bow shock radius units; 5) angle between magnetic and rotation axis of pulsar.

| Model | $\zeta$ | $\eta$ | a | $\alpha$ |
|---|---|---|---|---|
| r3-f | 0 | 1/4 | 3 | 45° |
| r3-b | $\pi/2$ | 1/4 | 3 | 45° |
| r3-fb | $\pi/4$ | 1/4 | 3 | 45° |
| r6-f | 0 | 1/25 | 6 | 45° |
| r6-b | $\pi/2$ | 1/25 | 6 | 45° |
| r6-fb | $\pi/4$ | 1/25 | 6 | 45° |
| r18-f | 0 | 1/289 | 18 | 45° |
| r18-b | $\pi/2$ | 1/289 | 18 | 45° |
| r18-fb | $\pi/4$ | 1/289 | 18 | 45° |

### 3.1 "Bird's-eye" view

In Fig. 3 we demonstrate "bird's-eye" view of different configurations. We explore various thrust ratios (we consider cases $\eta \leq 1$, so that the resulting shock structure wraps around the pulsar) and effects of different internal geometries. While for not very small thrust ratio of $\eta = 1/4$ (left column) the differences between different geometries are mild, for very small thrust ratios of $\eta = 1/25, 289$ we find qualitatively different behavior for different orientations and different cuts.

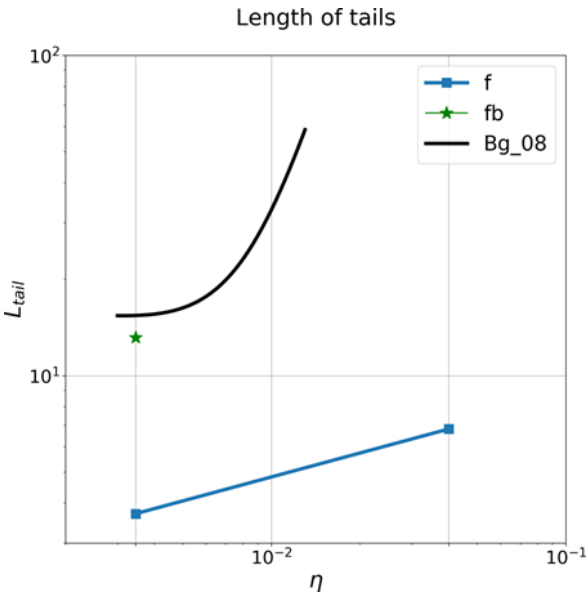

**Figure 7.** Dependence of length of tail $L_{\text{tail}}$ on the eta parameter $\eta$. The "Fris- bee" and "Frisbee-bullet" orientations were marked by 'squares' and 'stars' correspondingly. The measured parameters for the models are presented in the Tab.4. The black thick solid line denotes the result obtained in the work (Bogovalov et al. 2008).

There are two different shock wave structures that we obtained: opened-in-the-back and closed, qualitatively described in Fig. 2; results of simulations are in Figs. 4-5. Opened configurations have two branches laying along $x$-axis, which don't intersect each other. The upper situation in figure 4 may be only in "Frisbee-Bullet" orientation, and the lower situation may be in "Frisbee" and "Bullet" orientation. Otherwise, in closed configuration, two branches have the intersection point.

In figures 4 and 5 the cross indicates the location of the neutron star. The distance between the neutron star and the intersection point of the shock wave "head" and $x$-axis is denoted by $R_x$. If the shock wave head has the apex point that don't laying at the $x$-axis (see upper sketch in fig. 4), the corresponding distance along x-axis is denoted as $R_a$. Also, in the case of open areas, the values of the angles between the branches and $x$-axis $\theta_1$ and $\theta_2$ were calculated. The clockwise is positive direction and the counterclockwise is negative. In the opposite situation of closed zone, the following parameters are calculated: the length of the zone from apex point to the end of the tail L, the locations of the maximum points in perpendicular directions to the $x$-axis from the location of the neutron star ($x_+$ and $x_-$) and from the $x$-axis ($h_+$ and $h_-$). The corresponding values of the parameters are presented in tables 3 and 4.

In Figs. 6-7 we presented the dependence of the geometrical properties on the parameters of the simulations. In these figures, 'squares', 'triangles' and 'stars' denote "Frisbee", "Bullet" and "Frisbee-Bullet" orientation correspondingly. The $\eta$ parameter was chosen as the parameter of the $x$-axis, which has the connection with distance from neutron star to normal star:

$$\frac{r}{a} = \frac{\sqrt{\eta}}{1 + \sqrt{\eta}}, \quad (2)$$

One can see that wind zone expands with increasing of $\eta$ parameter. This result has the similarities with the paper Bogovalov et al. 2008. In figure 6 the black thick solid line means the result obtained by Bogovalov et al. 2008. But this dependence on the eta parameter has a gradient that is greater than when taking into account the magnetic field. On the other hand, not taking into account magnetic field by Bogovalov et al. 2008 leads to critical value of the $\eta$ parameter, which means that there are only opened configurations for greater values $\eta$. Figure 7 shows that closed configuration takes place when magnetic field has Frisbee or Frisbee-Bullet orientation, and wind zone can take smaller values of tail lengths compared to the result without taking into account the magnetic field.

In other words, for pulsar with the spin axis in the orbital plane, the length of the free wind zone can change from 1/4 orbital separation till $x \approx 3v_w/2\Omega\eta^{1/2} \sim 2T_{\text{orb},6}^{1/3}\eta^{1/2}$ (Barkov and Bosch-Ramon 2021). So photon-photon absorption can vary in one order of magnitude and lead to strong variation of gamma-ray signal during orbital period (see as example Khangulyan, Aharonian, and Bosch-Ramon 2008; Zabalza et al. 2013).

### 3.2 Dynamics of the magnetized wind

Next, we discuss the internal magnetic field structure of the flow. In figures 8-10 we show a complex plot of pressure (by color) which is projected on the CD surface and streamlines showing magnetic field strength by its thickness and color.

**Table 3.** Parameters of the shock waves - opened configuration

| Name | plane | $R_x$ | $R_a$ | $\theta_1$ | $\theta_2$ |
|---|---|---|---|---|---|
| r3-b | XZ | 0.1 | 0.2 | 0.37 | -0.36 |
|  | XY | 0.1 | 0.2 | 0.45 | -0.28 |
| r3-f | XZ | 0.6 | - | 0.10 | -0.09 |
|  | XY | 0.6 | - | 0.36 | -0.35 |
| r3-fb | XZ | 0.4 | - | 0.47 | -0.50 |
|  | XY | 0.4 | - | 0.35 | -0.30 |
| r6-b | XZ | 0.1 | 0.3 | 0.19 | -0.11 |
|  | XY | 0.1 | 0.3 | 0.18 | -0.09 |
| r6-fb | XZ | 0.5 | - | 0.37 | -0.30 |
|  | XY | 0.4 | - | 0.04 | -0.02 |
| r18-b | XZ | 0.2 | 0.4 | 0 | 0 |
|  | XY | 0.2 | 0.4 | 0 | 0 |

**Table 4.** Parameters of the shock waves - closed configuration

| Name | plane | $R_x$ | $h_+$ | $h_-$ | $x_+$ | $x_-$ | $L$ |
|---|---|---|---|---|---|---|---|
| r6-f | XZ | 0.68 | 0.68 | 0.67 | 1.19 | 1.22 | 6.82 |
|  | XY | 0.68 | 2.00 | 2.00 | 3.06 | 2.98 |  |
| r18-f | XZ | 0.78 | 0.60 | 0.65 | 1.09 | 0.87 | 3.71 |
|  | XY | 0.78 | 1.57 | 1.57 | 1.47 | 1.47 |  |
| r18-fb | XZ | 0.50 | 1.14 | 2.14 | 2.24 | 7.98 | 13.16 |
|  | XY | 0.41 | 1.22 | 1.22 | 0.40 | 0.42 |  |

We find that the shape of the CD changes from almost conical one ("Bullet" cases), transformed to flattened structure for the "Frisbee-Bullet" cases, and cross for "Frisbee/Cartwheel". The jet in the "Bullet" configurations pushed bow-shock closer to the normal star and magnetic field near CD head of the jet is significantly increased compare to "Frisbee-Bullet" and "Frisbee/Cartwheel" cases.

In figures 11-13 we present density and streamlines distribution, color indicates velocity magnitude for different winds thrust ration $\eta = 1/4$, $\eta = 1/25$, and $\eta = 1/289$ respectively. On all plots we see standard structure of the flow which consist of free stellar wind and pulsar wind separated by shocked pulsar wind and stellar wind. The magnetic field significantly affects the flow from the pulsar. In an equatorial plane wind is stronger (see Eq. 6) but in the polar direction hoop stress climate flow in the jet like structures.

The density plots indicate clearly the CD. The angle of the CD is dependent on the $\eta$ parameter and less sensitive to the pulsar orientation. The position of the pulsar wind shock, much more sensitive to the pulsar orientation. The higher $\eta$ parameter leads to smoother (more laminar) flow for all orientations, smaller $\eta$ leads to more turbulent flow. The "Bullet" configuration forms turbulent motion near CD, but near pulsar wind shock the flow is smooth. The "Cartwheel" or "Frisbee-Bullet" cases show strong turbulence in polar regions, especially if it is inclined onto normal star.

In figure 3 we presented the density cuts in different planes. Here we clearly see shocks in the stellar wind and in the pulsar wind. The orientation of the pulsar significantly affects the reverse shock geometry. The results in general are similar to results obtained in the works (Barkov, Lyutikov, and Khangulyan 2019; Barkov et al. 2019). For higher values of $\eta$ parameters, the flow become more expanding, but the general topology is preserved.

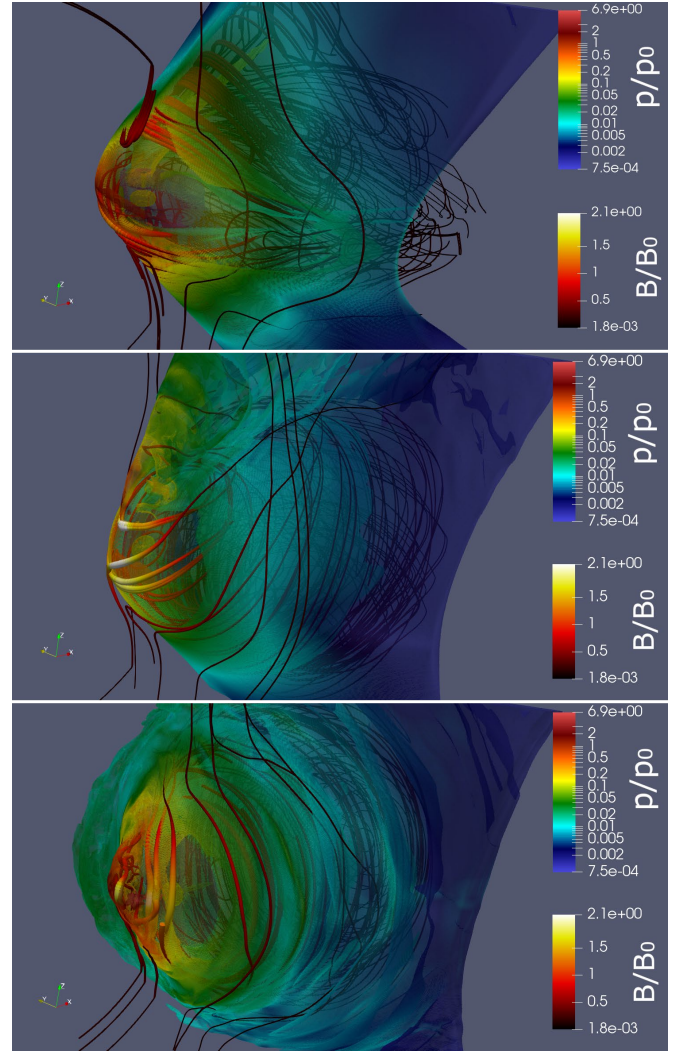

**Figure 8.** The pressure (color) and magnetic field lines (streamlines) for models with $\eta = 1/4$ r3-f (top panel), r3-fb (middle panel) and r3-b (bottom panel). We extrude the data at the contact discontinuity based on the jump of density and put half-transparent surface of pressure by color. The thickness of the streamlines indicates the strength of the magnetic field. The quantities $B_0$ and $p_0$ are means the units of magnetic field and pressure consequently ($B_0$ is equal to 52 G and $p_0$ is equal to 137 bar).

## 4. Radiation processes

We calculated the nonthermal emission of the pulsar in high-mass binary system, as a prototype, we chose LS 5039. The properties of the binary system used are shown in Table 5 (Casares et al. 2005). Thus, using the formula 1 one can find the spin-down luminosity of a neutron star (see Table 6). The bolometric luminosity of the system is dominated by optical emission of the normal massive star. On the other hand, nonthermal radiation processes are controlled by the synchrotron (SYN) and inverse Compton (IC) (see for details

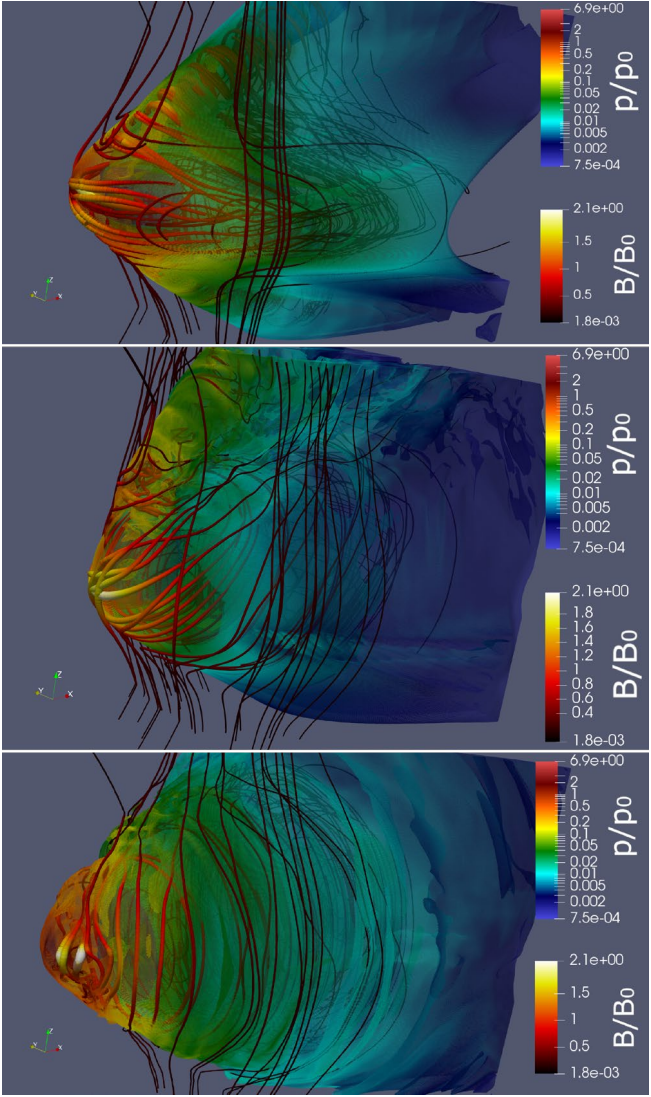

**Figure 9.** Same as Fig. 8 for models with $\eta = 1/25$ r6-f (top panel), r6-fb (middle panel) and r6-b (bottom panel). We extrude the data at the contact discontinuity based on the jump of density and put half-transparent surface of pressure by color. The thickness of the streamlines indicates the strength of the magnetic field. The quantities $B_0$ and $p_0$ are means the units of magnetic field and pressure consequently ($B_0$ is equal to 42 G and $p_0$ is equal to 88 bar).

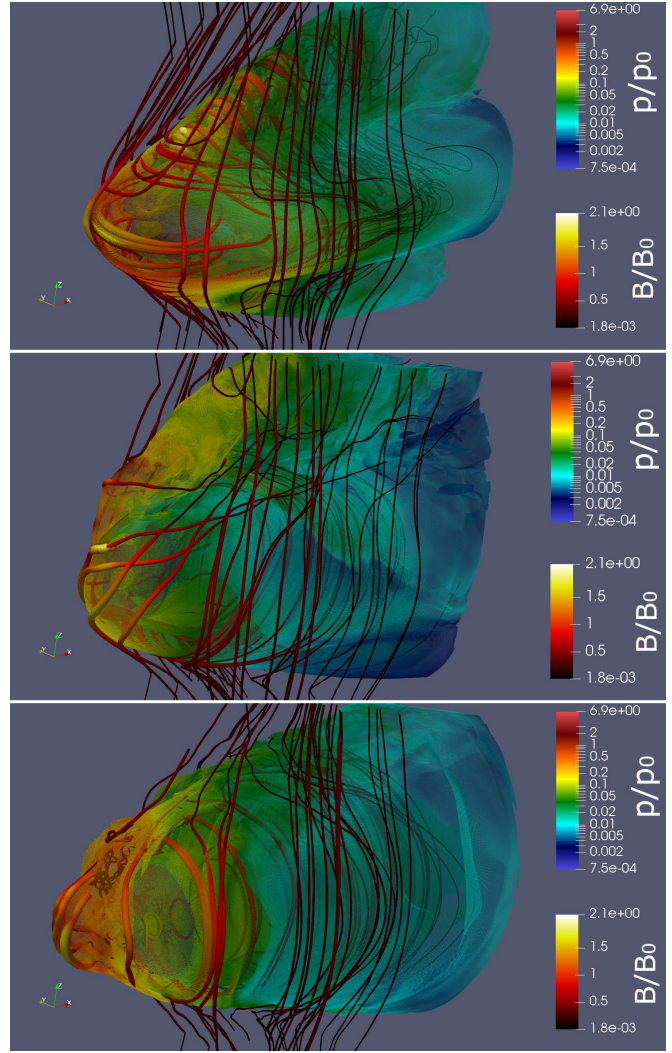

**Figure 10.** Same as Fig. 8 for models with $\eta = 1/289$ r18-f (top panel), r18-fb (middle panel) and r18-b (bottom panel). We extrude the data at the contact discontinuity based on the jump of density and put half-transparent surface of pressure by color. The thickness of the streamlines indicates the strength of the magnetic field. The quantities $B_0$ and $p_0$ are means the units of magnetic field and pressure consequently ($B_0$ is equal to 37 G and $p_0$ is equal to 68 bar).

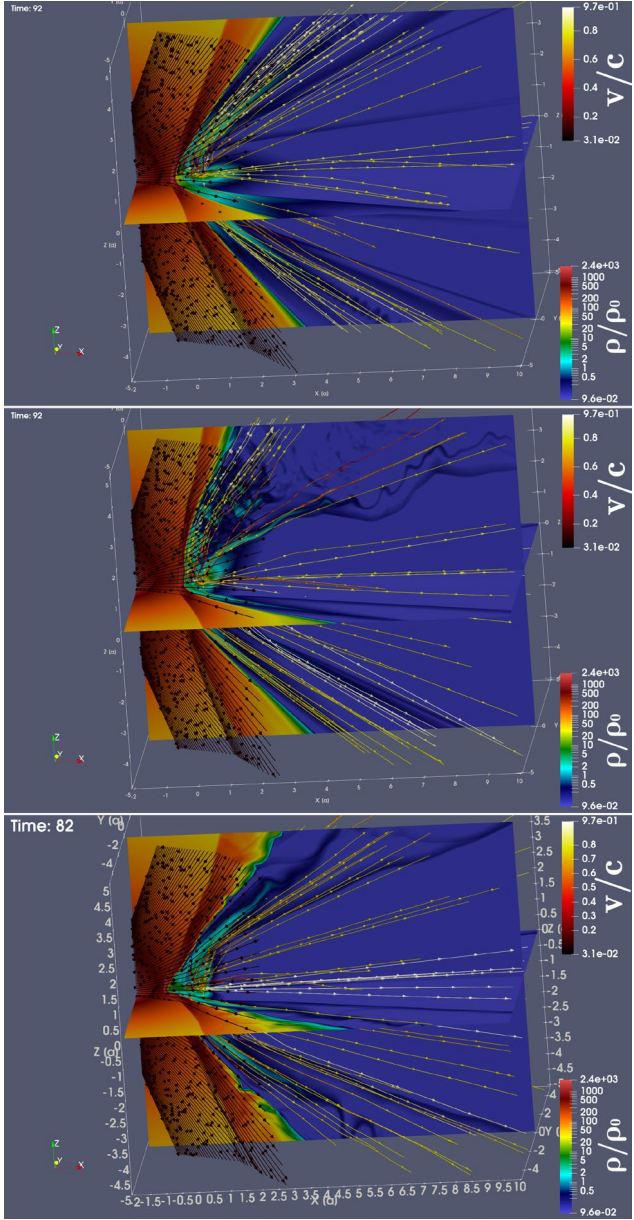

**Figure 11.** The density (color) and velocity distribution (streamlines) for the models with $\eta = 1/4$ r3-f (top panel), r3-fb (middle panel) and r3-b (bottom panel). The quantities $\rho_0$ and $c$ means the units of density and speed consequently ($\rho_0$ is equal to $6.33 \times 10^{-18} gr/cm^3$ and $c$ is the speed of light).

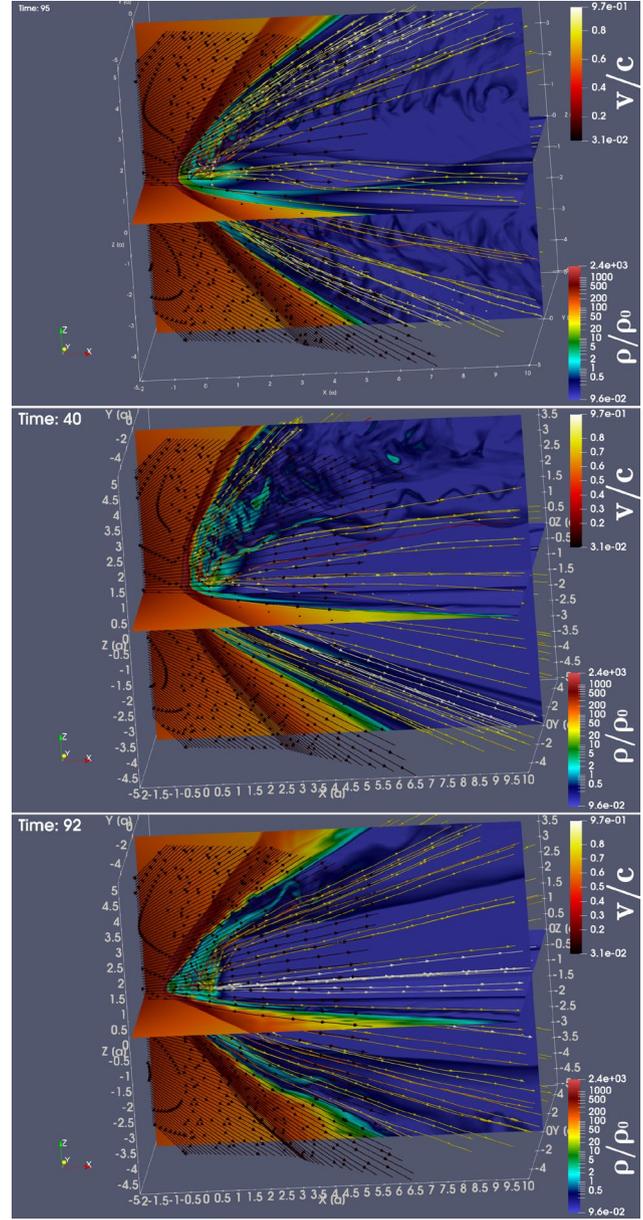

**Figure 12.** The density (color) and velocity distribution (streamlines) for the models with $\eta = 1/25$ r6-f (top panel), r6-fb (middle panel) and r6-b (bottom panel). The quantities $\rho_0$ and $c$ means the units of density and speed consequently ($\rho_0$ is equal to $4.05 \times 10^{-18} gr/cm^3$ and $c$ is the speed of light).

Khangulyan, Aharonian, and Bosch-Ramon 2008), and $\gamma - \gamma$ absorption has a significant effect on hard radiation and has to be taken into account as well. Our emissivity model is similar to the described in the (Bosch-Ramon and Khangulyan 2009; Khangulyan, Aharonian, and Kelner 2014), details are presented in the Appendix 2.

To evaluate the gamma-gamma absorption process, optical thickness $\tau = \int n\sigma dl$ was calculated assuming maximum cross-section for two-photon pair production $\sigma$ ($\sigma_T/5$) was selected. Optical thickness $\tau_0$ was calculated for one point located at a unit distance from a massive star, then the remaining values along one straight line were calculated using the scaling property of the optical thickness (see formula (12) Khangulyan, Aharonian, and Bosch-Ramon 2008) as:

$$\tau = \tau_0 \frac{d}{l}, \quad (3)$$

where $\tau_0$ is evaluated optical thickness on the distance $d = 1$ from star, $l$ is the distance for which we evaluate thickness $\tau$.

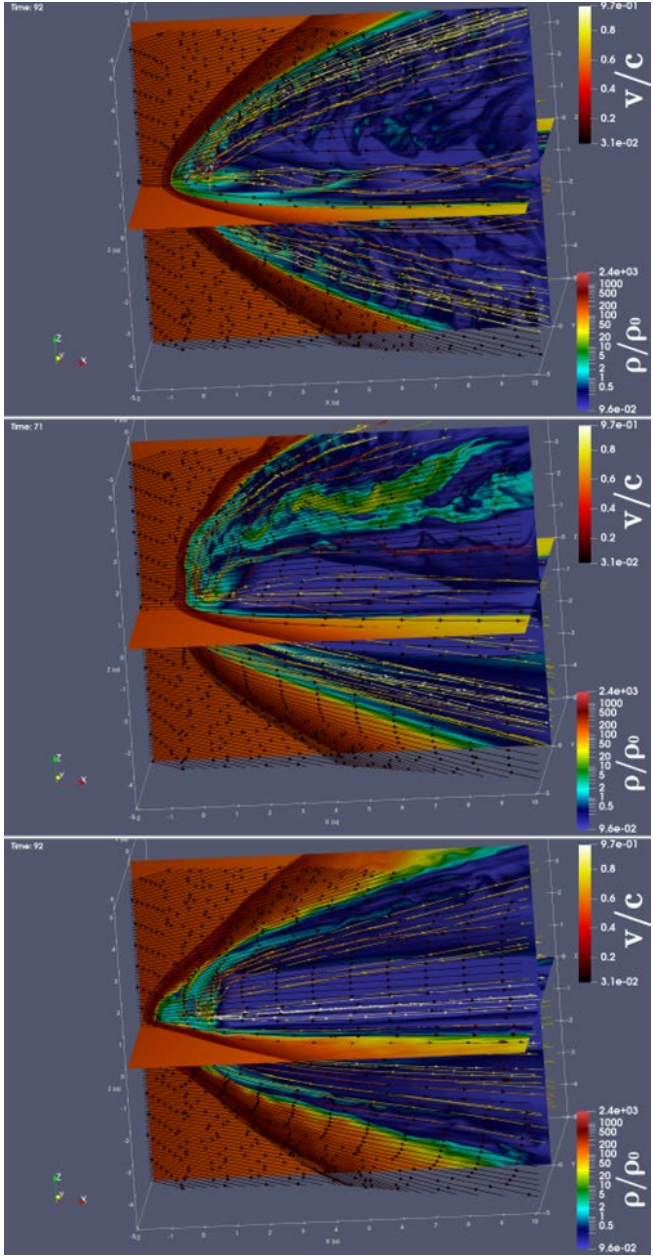

**Figure 13.** The density (color) and velocity distribution (streamlines) for the models with $\eta = 1/289$ r18-f (top panel), r18-fb (middle panel) and r18-b (bottom panel). The quantities $\rho_0$ and $c$ means the units of density and speed consequently ($\rho_0$ is equal to $3.15 \times 10^{-18} gr/cm^3$ and $c$ is the speed of light)

**Table 5.** Parameters of the binary system LS 5039

| Parameter | Value |
|---|---|
| Star luminosity [erg $s^{-1}$] | $7 \times 10^{38}$ |
| Star temperature [K] | $3.9 \times 10^4$ |
| Stellar radius [$R_\odot$] | 9.3 |
| Orbital semi-major axis [cm] | $2.1 \times 10^{12}$ |
| Eccentricity | 0.35 |
| Wind velocity [cm $s^{-1}$] | $2 \times 10^8$ |
| Mass loss rate [$M_\odot$ $yr^{-1}$] | $5 \times 10^{-7}$ |

**Table 6.** Pulsar luminosity on the model

| Name | $\eta$ | Lsd, erg/s |
|---|---|---|
| r 3 | 1/4 | $4, 5 \times 10^{37}$ |
| r6 | 1/25 | $7, 2 \times 10^{36}$ |
| r18 | 1/289 | $6, 3 \times 10^{35}$ |

The emissivity for the IC and SYN processes was calculated using the formula $I'_{rad} \propto \int (u_{cell}/t_{rad})d\ell$, where $\ell$ is the length element along the line of sight (see Barkov and Bosch-Ramon 2018). We take into account the power-low distribution of electrons with spectra $n = A\epsilon^{-\alpha}$ there an $\alpha$ = 2. The resulting value describes the emissivity of the system in its own frame of reference. The Doppler factor $\delta$ (see Appendix 2) is used to convert the emissivity from the co-moving reference frame $I'_\nu$ to a laboratory reference frame $I_\nu$. Thus, the emissivity for $\alpha = 2$ is converted by the formula.

$$I_\nu = \delta^{2.5} I'_\nu, \quad (4)$$

In particular, in the case of the IC process, gamma absorption of gamma radiation was taken into account by multiplying by an additional factor:

$$I_{IC} = I_\nu exp(-\tau), \quad (5)$$

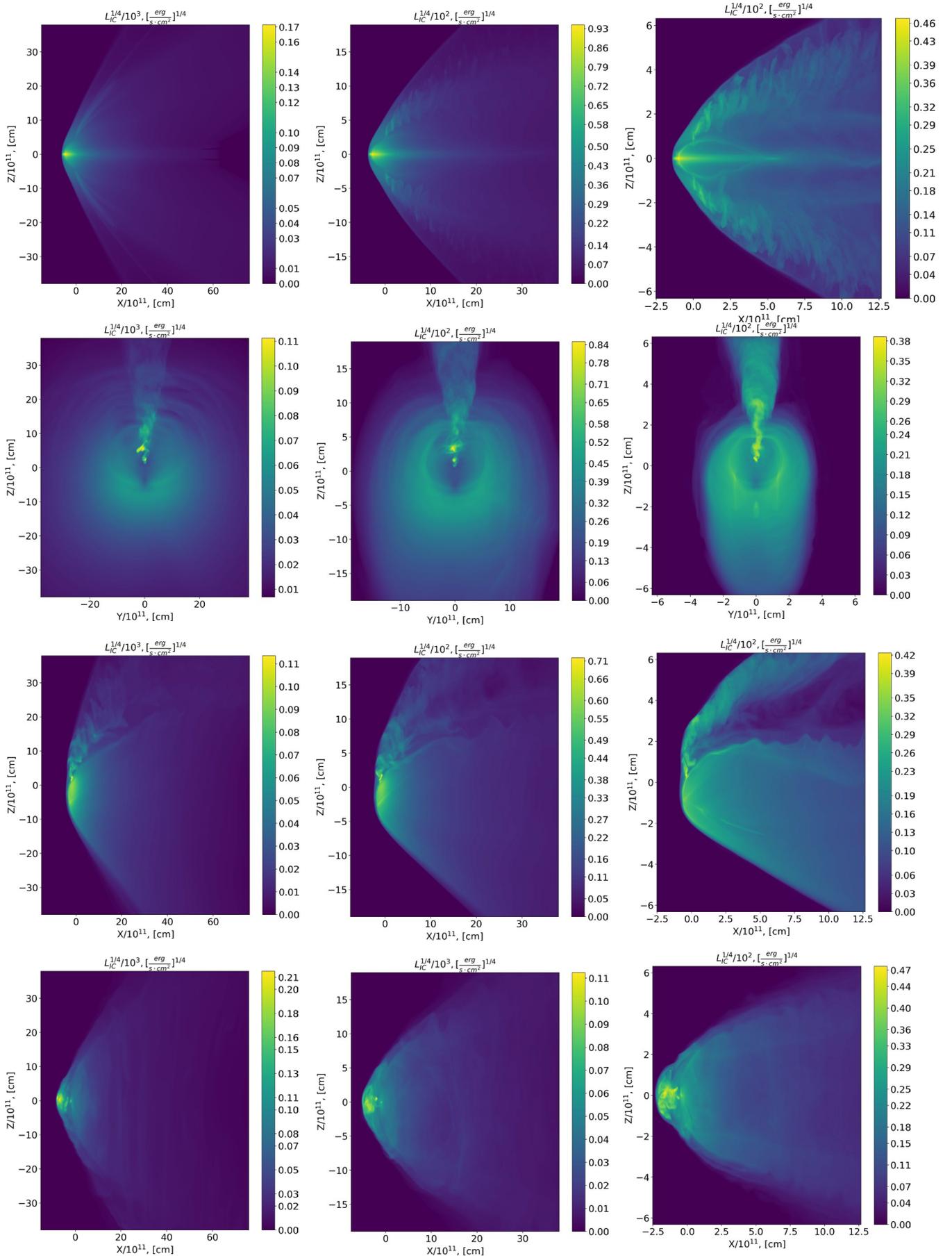

**Figure 14.** Emissivity maps for IC process with $\eta = 1/4$ (left column), $\eta = 1/25$ (central column), and $\eta = 1/289$ (right column). At the first row "Frisbee" for line of sight along Y-axis, the "Frisbee-Bullet" for lines of sight along X-axis and Y-axis on the second and third rows respectively, and at the fourth row "Bullet" for line of sight along Y-axis.

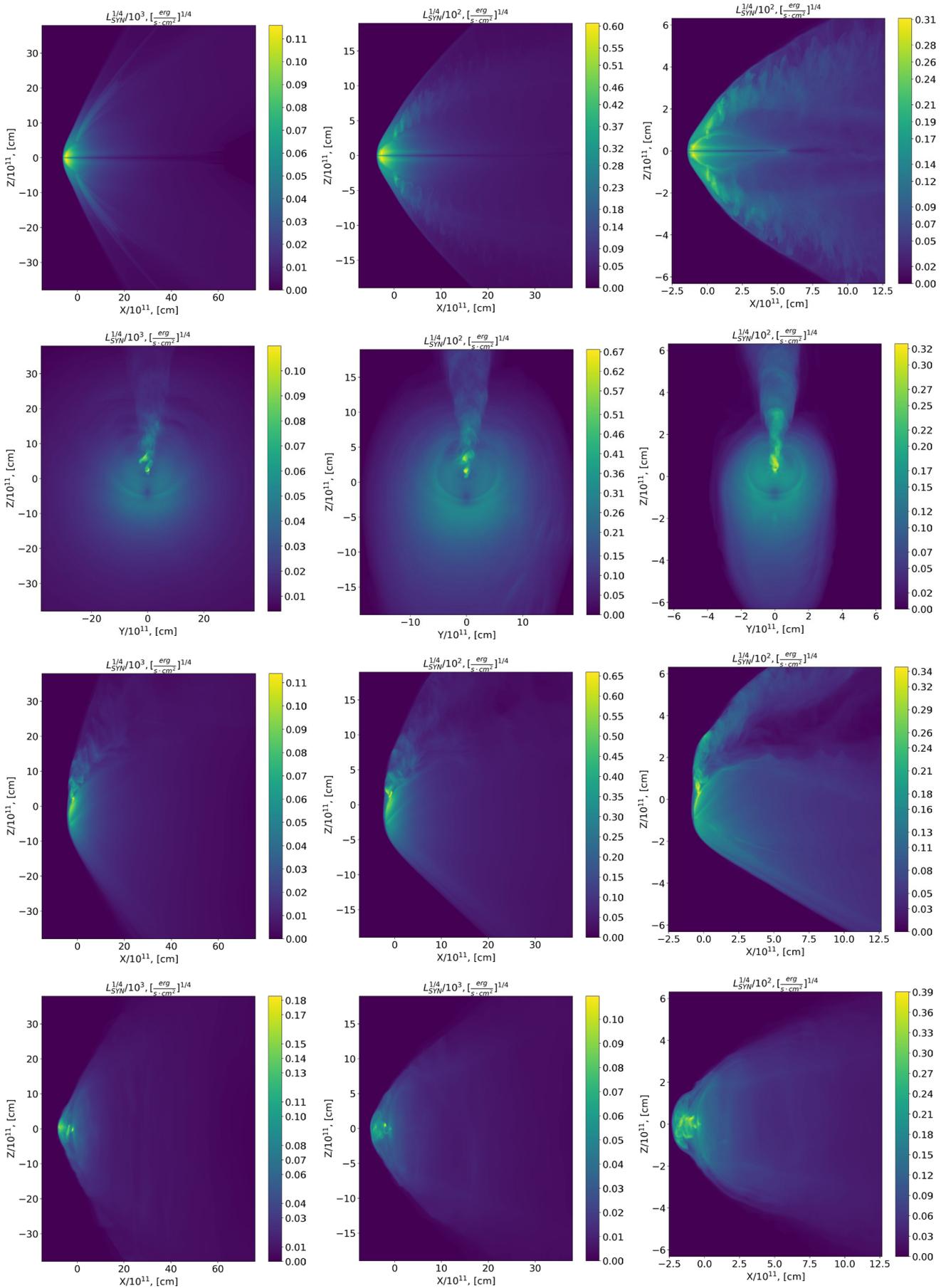

**Figure 15.** Emissivity maps for SYN process with $\eta = 1/4$ (left column), $\eta = 1/25$ (central column), and $\eta = 1/289$ (right column). At the first row "Frisbee" for line of sight along the Y axis, the "Frisbee-Bullet" for lines of sight along the X axis and the Y axis on the second and third rows, respectively, and at the fourth row "Bullet" for line of sight along Y-axis.

where $\tau$ is the optical thickness (see formula (3)).

We calculated emissivity maps which describe the system emission along the X, Y, Z axes (lines of sight, see fig. 14 and 15). The flow from the neutron star which form jets and the front bow shock are the brightest details. Next one is the pulsar shocked wind at larger distance from the pulsar which form quite bright background. We should note, the bright element of the flow is the back shock and pulsar equatorial region, which in total can give luminosity values comparable to the luminosity of individual parts of the front shock.

The luminosity maps depending on the polar ($\theta$) and azimuthal ($\varphi$) angles are presented on (fig. 16), for sake of simplicity here and below we set eccentricity $\epsilon = 0$. Further, other possible directions of the lines of sight to this system were considered, and the obtained emissivity was summed up over the entire surface, that is, the luminosity of the system for this line of sight was obtained. For visualization of the results obtained, lines of luminosity levels were plotted according to the angles of view (Fig. 16). The polar angle of $\theta$ is calculated from the orbital plane and takes values from $\pi/2$ (along the selected Z axis) to $-\pi/2$ (opposite to the Z axis). The azimuthal angle $\phi$ takes values from zero (X-axis) to $2\pi$ (also X-axis), increasing its value when rotating counterclockwise when looking at the system from above. It can be seen that for the IC process, the luminosity has its minimum value for the angles $\theta = 0$ and $\varphi = 0$ ($2\pi$), which is a consequence of the fact that a massive star takes place in the path of the line of sight.

It is noticeable that the "Frisbee" and "Frisbee-Bullet" configurations form symmetrical light curves, but in the "Bullet" configurations due to the stochastic influence of wind from a massive star, the jet deviates from the original direction (the line connecting the neutron and massive stars) backward, but by an arbitrarily chosen side, which violate the symmetry in the flow and light curves. This can be seen in the fifth and sixth rows of the figure 16. If the jet deviation did not occur, then bright spots in these drawings would be observed at the positions $\theta = 0$, $\varphi = 0$ and $\theta = 0$, $\varphi = 2\pi$.

As mentioned above, for the "Frisbee" configuration, the polar angle $\theta$ is responsible for the angle of inclination of the binary system, and the azimuth angle $\varphi$ is responsible for the orbital angle of rotation of the neutron star. Based on the results obtained (see figure 16), figures of the dependence of the luminosity of the binary system on the orbital angle ($\varphi$) in the "Frisbee" configuration for 5 different angles of inclination were constructed: $0°$, $30°$, $45°$, $60°$, and $90°$ (see figures 17). The angle of $\varphi$ varies from $-\pi$ to $3\pi$, describing two periods of rotation of a neutron star. From the constructed light curves the luminosity of the system does not change for SYN and IC radiation for an angle of inclination equal to $90°$. This is due to the fact that the binary system does not change its spatial position for the observer. The steepest figures are obtained when the angle of inclination is $0°$ (the binary system is located with an edge in relation to the observer). But the minimum of IC radiation is observed at the points where the neutron star is located behind a massive star ($\varphi = 0$ and $\varphi = 2\pi$), which is associated with photon-photon absorption.

If the observer's line of sight changed only with a change in the angle of $\varphi$, then the luminosity of the system may also differ for the upper and lower parts. This can be seen from the figure 16. For "Frisbee" and "Frisbee-Bullet" configurations, such flights in the upper part ($\theta$ from 0 to $\pi/2$) would almost not differ from the flight in the symmetrical lower part of the system ($\theta$ from 0 to $-\pi/2$, respectively). But at the same time, the flyby from above and from below in the same area (symmetrical points relative to the equatorial plane) are significantly different. So, for example, for the "Frisbee-Bullet" configuration, this difference can reach 100% for SYN and 120% for the IC process. In case "Bullet", the deviation of the jets noticeably affects the luminosity maps (fifth and sixth rows in the figure 16), which also affects the flight around the system in the upper and lower parts. It can be seen that with a lower value of the $\eta$ parameter ($\eta = 1/289$), the picture remains symmetrical with respect to the orbital plane ($\theta = 0$), but for large values of the $\eta$ parameter ($\eta = 1/4$), the difference in the luminosity of the upper and lower parts is noticeable.

Although the described configurations have a static appearance for a certain orientation of the spin of a neutron star, it is possible to construct the light curves of the system under consideration from the obtained results. All these configurations indicate different locations of the neutron star relative to the massive star during rotation. Knowing the angular velocity $\omega$ of the orbital motion of a neutron star and using a map of luminosity (fig. 16) with certain line of sight (which is described by angles $\theta$ and $\varphi$), it is possible to determine the time from the formula $t = \alpha/\omega$, where the angle $\alpha$ (in radians) is the phase of the orbital rotation.

If the spin of the neutron star is perpendicular to the orbital plane ("Frisbee"), then this orbital angle $\alpha$ will be the angle $\varphi$ from figure 16, and the light curve coincides with luminosity maps with fixed value of angle $\theta$. Thus, to construct the light curve in the case of the Frisbee configuration, it is sufficient to select a certain angle of view $\theta$ for rows five and six of figure 16 and to obtain the dependence of luminosity on the orbital angle $\varphi$. The light curves for the "Frisbee" configuration are shown in Fig. 17. If the spin lies in the orbital plane, there will be a change of configurations "Bullet"-"Frisbee-Bullet"- "Frisbee"-"Frisbee-Bullet"-"Bullet" every half of the period of movement around a massive star; therefore, in addition to the angle of $\varphi$, one need to use different luminosity maps (different rows of fig. 16). In this case, the configurations "Bullet" and "Frisbee" occur every half of the period, which corresponds to a change in the value of angle $\varphi_0$ to angle $\varphi_0 + \pi$, while the configuration "Frisbee-Bullet" is obtained every quarter of the period, which corresponds to a change in angle $\varphi_0$ to angle $\varphi_0 + \pi/2$. Figure 18 shows these light curves for a fixed distance between neutron and massive stars ($\eta = 1/4$). Eight different configurations of a neutron star (8 points on the graph) were used for plotting, which were connected using linear interpolation. The position when the neutron star is located in the infer-far conjunction in front of the massive star is taken as the beginning of the countdown. The figures differ in the initial configuration - the possible direction of the spin of the neutron star relative to the orbital moment. The time is normalized for the orbital motion period $P_s$, which for the LS5039 system is $P_s$ = 3.9$d$. It can be seen what, in some situations, the light curve for this system can change 8–10 times for a period.

On the figure 19 As we can see IC and synchrotron typical light curves behave in opposite directions. Maximum in the synchrotron corresponds to minimum in the IC and corresponds to inferior conjunction. We plot theoretical light curves for Synchrotron and IC emission on Fig. 20 and observed X-ray energy band light curve for LS 5039 (Yoneda et al. 2023). Nether the less LS 5039 system has small eccentricity, the peaks are reached in the inferior conjunction in both cases.

## 5. Discussion and Conclusion

In this work, we perform numerical relativistic MHD simulations of the wind-wind interaction in gamma-ray binaries. Qualitatively, our results connect previous hydrodynamical simulations of the wind-wind interaction of Bogovalov et al. 2008 and magnetized pulsar wind - ISM cases (Bogovalov et al. 2019; Barkov, Lyutikov, and Khangulyan 2019).

Our 3D simulations demonstrate that the wind-wind interaction in the gamma-ray binaries is different in many respects both from the fluid case and the pulsar wind - ISM cases. We find several features that are specific to 3D (R)MHD simulations that were naturally missed in the previous 2D RMHD work (see Bogovalov et al. 2012; Bogovalov et al. 2019). The two-dimensional simulations mostly investigated the dependence of the flow shape on the thrust ratio $\eta$ (Bogovalov et al. 2008). As we demonstrate here, the geometry of the pulsar wind is most important.

If compared with the wind-ISM interaction (Bogovalov et al. 2019), the main difference is the structure of the tail region of the PWN - in the case of planar ISM flow the pulsar wind always terminates at a tail ward Mach disk, while in the wind-wind case the pulsar wind may extend to large distances (of the order of several orbital separation). For example, Fig. 3, right column, shows that for a fixed $\eta$ the configuration changes from open to closed depending on the geometry of the flow.

In figure 6 we show the evolution of the RS opening angle $\theta$ depending on the parameter $\eta$ and the pulsar orientation angle $\zeta$. Our dependence on $\eta$ is generally consistent with the fluid case (Bogovalov et al. 2008). For 2D MHD simulations Bogovalov et al. 2019, found significant decrease of the opening angle with increasing magnetization of the pulsar wind. For $\eta = 0.3$ and $\sigma = 0$ the $\theta \approx 1$ then for $\sigma = 0.8$ the $\theta \approx 0.35$. Such the value is close to our "Frisbee-Bullet" orientation case. On the other hand, the bullet orientation trends to significantly wider RS angle, but "Frisbee/Cartwheel" in opposite trends to narrower opening angle, especially in directions in between pulsar spin axis and its equatorial plane. The jet in normal star direction in bullet configuration pushes away CD and did not allow the RS zone to close up to $\eta > 1/300$, which is significantly larger compared to the non-magnetized case $\eta > 1/50$.

Similar tendencies are observed in the closed configurations of pulsar wind (see Fig. 7). The non-magnetized axisymmetric wind forms longer tail compare to "Frisbee/Cartwheel" or "Frisbee-Bullet" configuration, perhaps similar to strongly magnetized wind. The bullet configuration did not form a closed configuration for our setup. In bullet configuration, jet-like structures push away the FS and reflect stellar wind like umbrella, so it decreases the wind gas pressure in the tail.

Our simulations do not take the dynamics effects of the orbital motion into account - only as a static sequence of different geometries. We can interpret our results as different individual binary systems with various *instantaneous* orientation of spin axis respect to orbital plane and with respect to the direction towards the companion: (i) if the pulsar spin axis is normal to the orbital plane, we have "Frisbee" configuration all the time; (ii) if the pulsar spin axis lay in the orbital plane, we have "Cartwheel" — "Frisbee-Bullet" — "Bullet" — "Frisbee- Bullet" — "Cartwheel" every half of the orbit; (iii) if the pulsar
spin axis angled on 45º to the orbital plane we have transition "Frisbee" — "Frisbee-Bullet" — "Frisbee" every half of the orbit. Thus, taking into account the anisotropic pressure effects of the magnetic field in the pulsar wind leads not only to formation of less compressible outflow in the shocked region, but formation of significantly different flow shape.

We demonstrate that the orientation of the pulsar's spin axis plays a significant role in the geometry of the shocked pulsar wind. For example, if the pulsar rotation axis lay in the orbital plane, then *the length of the free pulsar wind zone can change during 1/4 of the orbit almost one order of magnitude*. This can significantly change the emission for the distant observer. We hypothesize that this can be a crucial factor for TeV emission in gamma-ray binaries that suffer from strong photon-photon absorption, up to factor 10 (Khangulyan, Aharonian, and Bosch-Ramon 2008; Zabalza et al. 2013). Change in the absorption will drastically affect the light curves as well as spectra of such systems.

In the second stage, which is much more numerically challenging, we plan to conduct full 3D relativistic MHD simulations of the orbital motion of the pulsar. Here are the effects of the Coriolis and centrifugal forces becomes important. In the case of seemingly open tail configurations, eventually, at a distance comparable to the orbital separation, the wind zone will be closed due to Coriolis effects (see Bosch-Ramon and Barkov 2011; V. Bosch-Ramon et al. 2012). Even at the present setup, modelled light curves show feasible behaviour, see Fig. 20. The modelling of the orbital effects should make light curves more asymmetric and closer to observed one. Also, the inclination of pulsar's rotation axis can be significant, the "Frisbee-Bullet" configuration, that can lead to more smooth and asymmetric light curve formation.

**Acknowledgement**
The simulations were performed on CFCA XC50 cluster of National Astronomical Observatory of Japan (NAOJ) and RIKEN HOKUSAI Bigwaterfall.

**Funding Statement**   This research was supported by grant NASA 80NSSC20K1534 and 23-22-00385 of the Russian Sci-

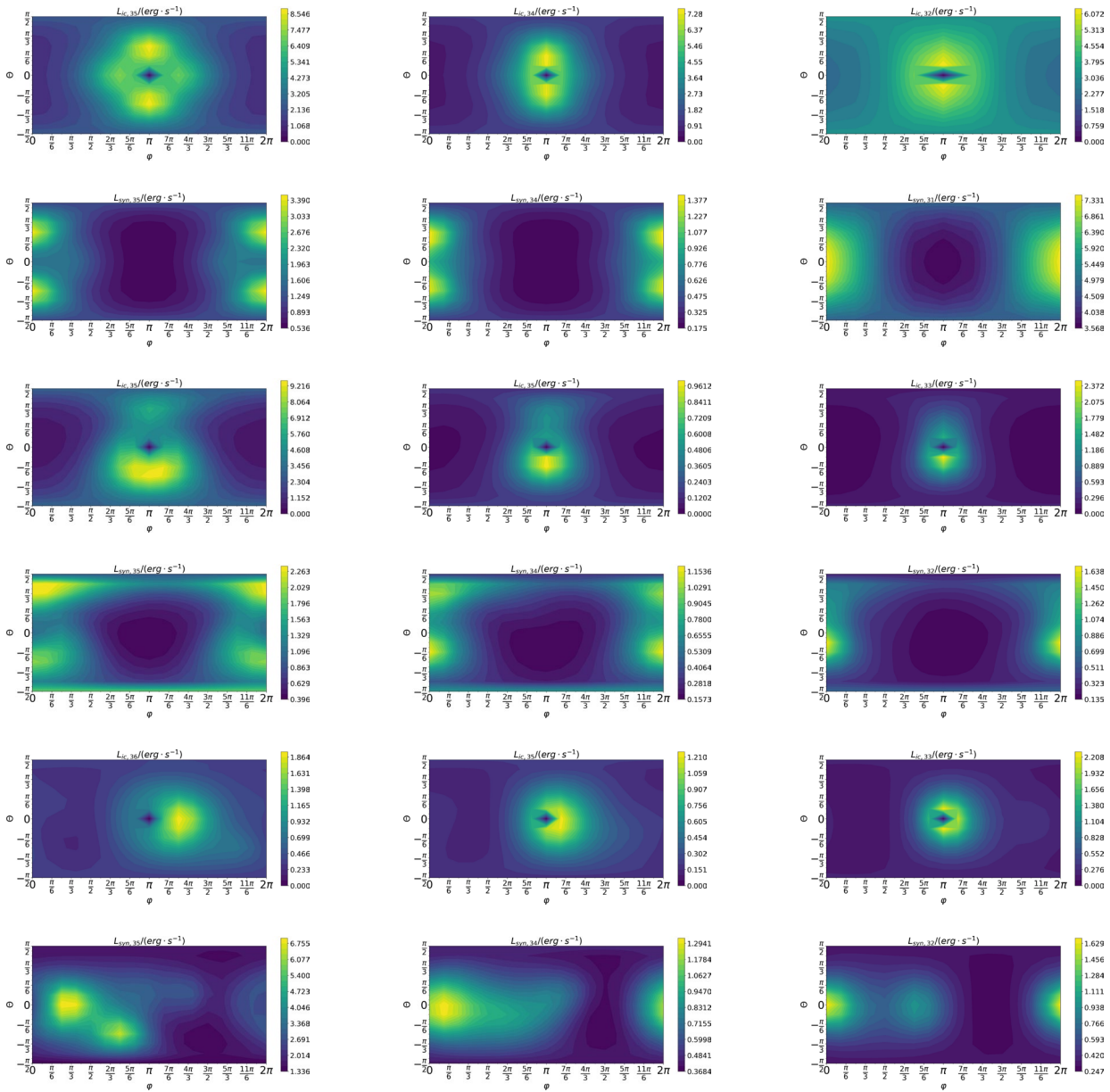

**Figure 16.** Luminosity maps depending on the polar ($\theta$) and azimuthal ($\varphi$) angles for the models with $\eta = 1/4$ (left column), $\eta = 1/25$ (central column), and $\eta = 1/289$ (right column). At the first row "Frisbee" IC process and the second row SYN process, the "Frisbee-Bullet" IC process and SYN process on the third and fourth rows respectively, and the "Bullet" IC process and SYN process on the fifth and sixth rows respectively.

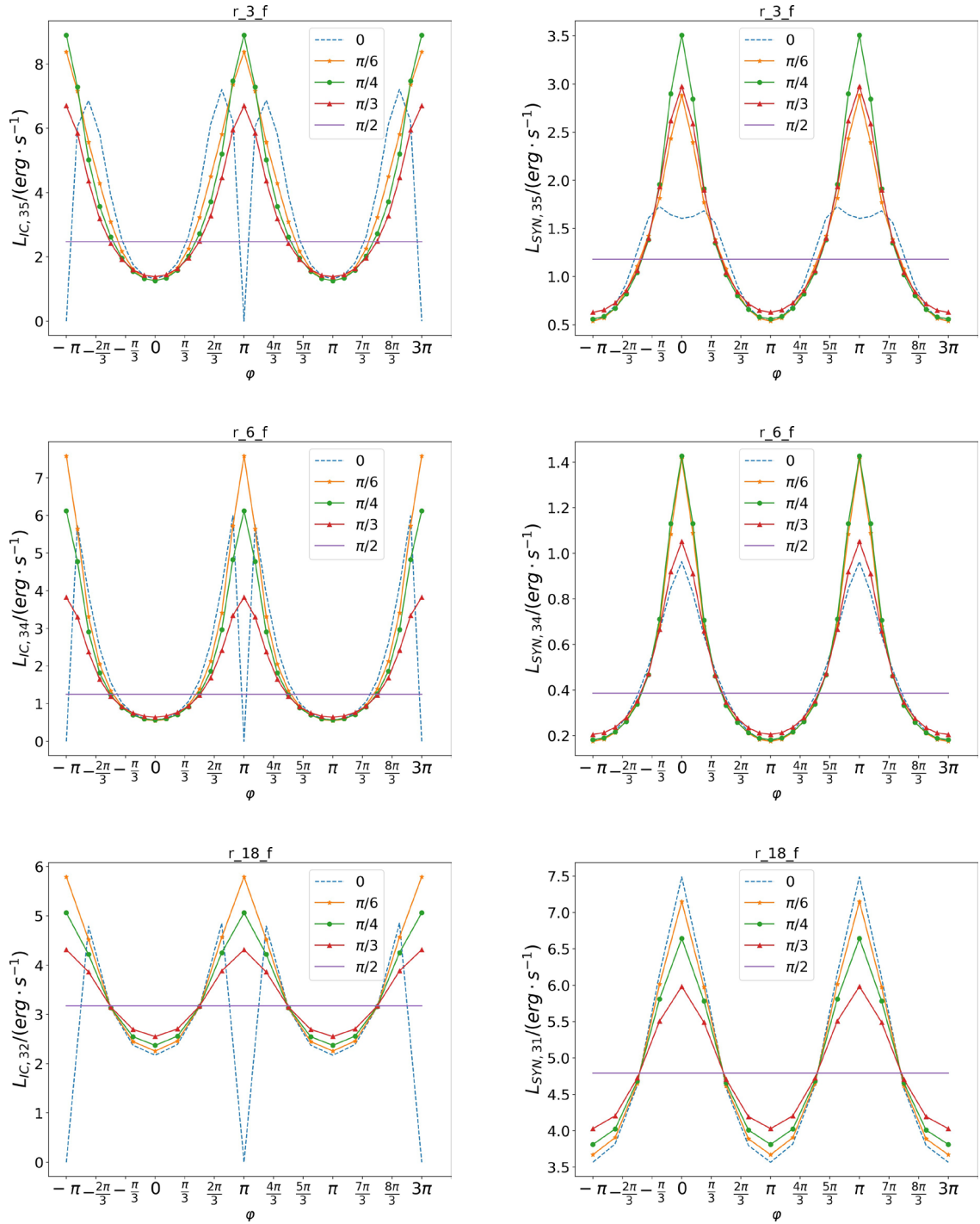

**Figure 17.** The dependence of luminosity on the orbital angle of a neutron star for a certain type of radiation (IC and SIN) for six different angles of $\theta$: $0$, $\pi/6$, $\pi/4$, $\pi/3$ and $\pi/2$. The orbital angle $\varphi$ varies from $-\pi$ to $3\pi$, where the angle $\varphi = 0$ means the direction to the observer.

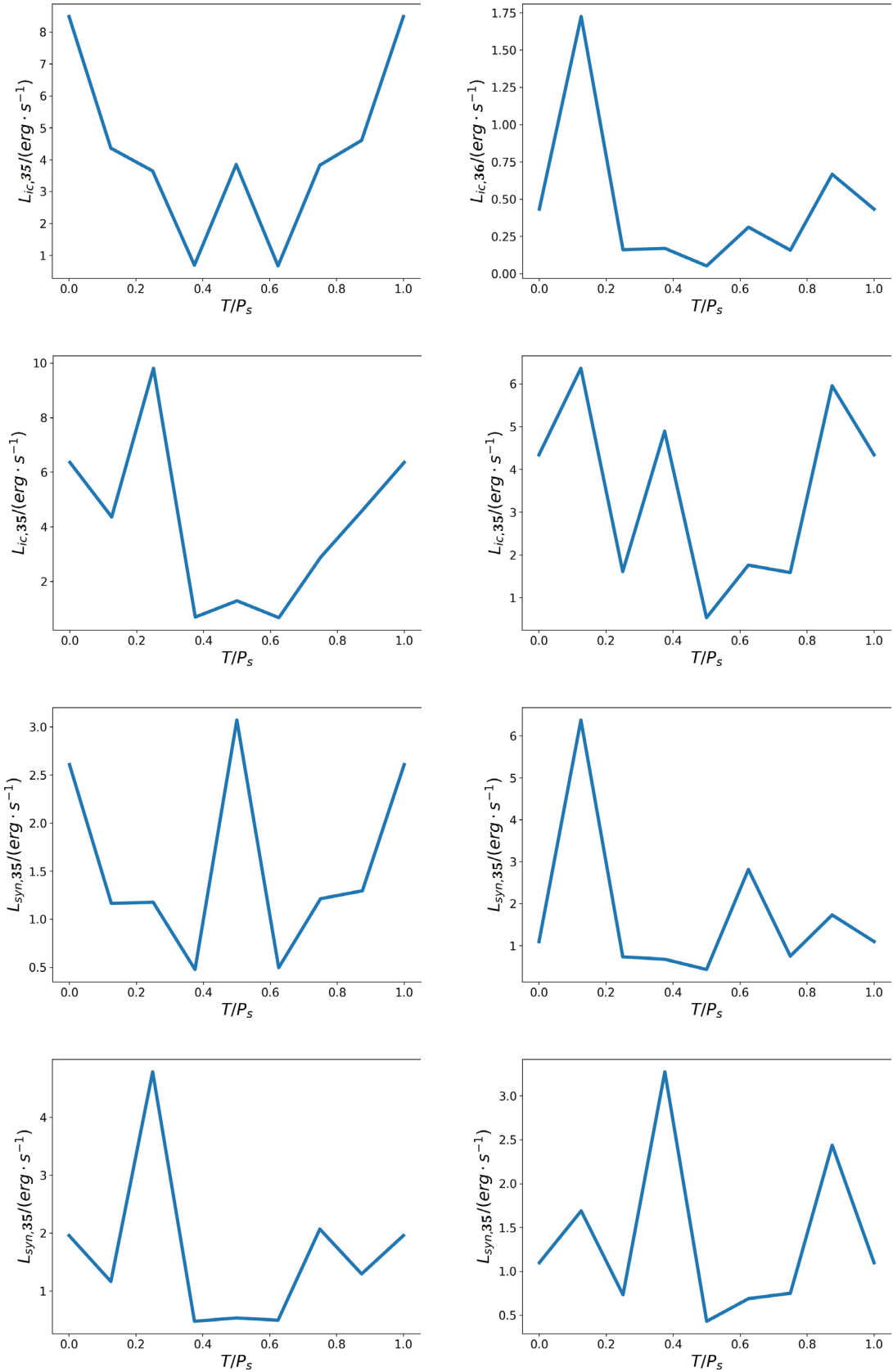

**Figure 18.** Light curves of equatorial plane of the processes IC (first and second rows) and SYN (third and fourth rows) for $\eta = 1/4$. The time is normalized to the period of the orbital motion of the neutron star. The configuration changes as "Bullet"-"Frisbee-Bullet"-"Frisbee"-"Frisbee-Bullet"-"Bullet" every half period. In the first and third rows, the initial configurations are "Bullet"-"Frisbee-Bullet" for the left column or "Frisbee-Bullet"-"Bullet" for the right column. In the second and fourth rows, the initial configurations are "Frisbee"-"Frisbee-Bullet" for left column or "Frisbee-Bullet"-"Frisbee" for right column. At the moment $T = 0$, the neutron star is located in an infer-far conjunction between the massive star and observer.

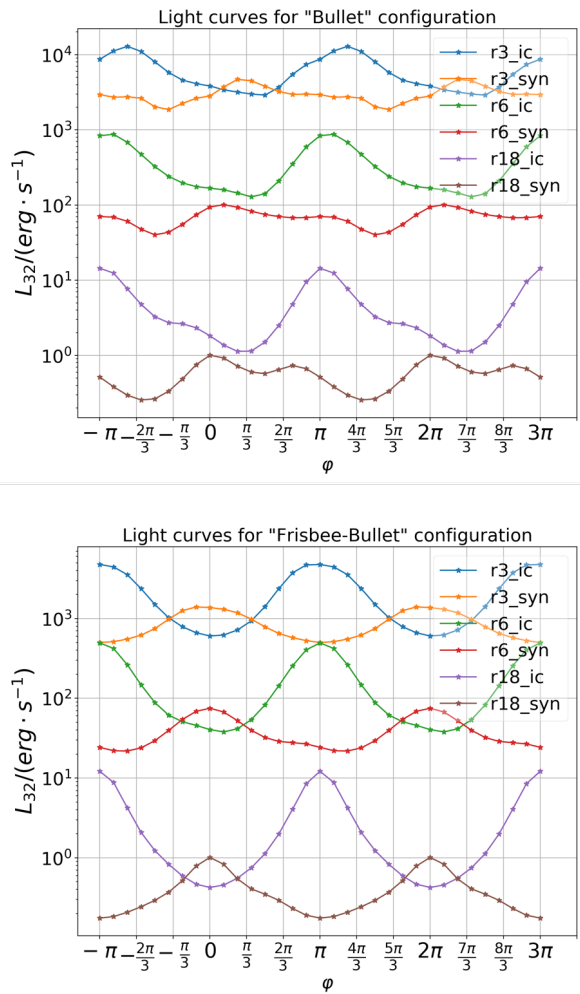

**Figure 19.** Light curves of $\theta = \pi/6$ for different types of configurations: "Bullet" (top panel) and "Frisbee-Bullet" (bottom panel). The figure has six light curves, which has changed the $\eta$-parameter ($\eta = 1/4$ for r3, $\eta = 1/25$ for r6, and $\eta = 1/289$ for r18) and types of radiation (SYN or IC). The orbital angle $\varphi$ changes from $-\pi$ to $3\pi$. The $\varphi = 0$ corresponds to the neutron star inferior conjunction.

**References**

Abdo, A. A., M. Ackermann, M. Ajello, A. Allafort, J. Ballet, G. Barbiellini, D. Bastieri, et al. 2011. Discovery of High-energy Gamma-ray Emission from the Binary System PSR B1259-63/LS 2883 around Periastron with Fermi. *ApJ* 736, no. 1 (July): L11. https://doi.org/10.1088/2041-8205/736/1/L11. arXiv: 1103.4108 [astro-ph.HE].

Barkov, Maxim V., and Valenti Bosch-Ramon. 2016. The origin of the X-ray-emitting object moving away from PSR B1259-63. *MNRAS* 456, no. 1 (February): L64–L68. https://doi.org/10.1093/mnrasl/slv171. arXiv: 1510.07764 [astro-ph.HE].

———. 2018. A hydrodynamics-informed, radiation model for HESS J0632 + 057 from radio to gamma-rays. *MNRAS* 479, no. 1 (September): 1320–1326. https://doi.org/10.1093/mnras/sty1661. arXiv: 1806.05629 [astro-ph.HE].

———. 2021. The Major Role of Eccentricity in the Evolution of Colliding Pulsar-Stellar Winds. *Universe* 7, no. 8 (July): 277. https://doi.org/10.3390/universe7080277. arXiv: 2204.07494 [astro-ph.HE].

Barkov, Maxim V., and Dmitry V. Khangulyan. 2012. Direct wind accretion and jet launch in binary systems. *MNRAS* 421, no. 2 (April): 1351–1359. https://doi.org/10.1111/j.1365-2966.2012.20403.x. arXiv: 1109.5810 [astro-ph.HE].

Barkov, Maxim V., Maxim Lyutikov, and Dmitry Khangulyan. 2019. 3D dynamics and morphology of bow-shock pulsar wind nebulae. *MNRAS* 484, no. 4 (April): 4760–4784. https://doi.org/10.1093/mnras/stz213. arXiv: 1804.07327 [astro-ph.HE].

———. 2020. Fast-moving pulsars as probes of interstellar medium. *MNRAS* 497, no. 3 (September): 2605–2615. https://doi.org/10.1093/mnras/staa1601. arXiv: 2002.12111 [astro-ph.HE].

Barkov, Maxim V., Maxim Lyutikov, Noel Klingler, and Pol Bordas. 2019. Kinetic 'jets' from fast-moving pulsars. *MNRAS* 485, no. 2 (May): 2041–2053. https://doi.org/10.1093/mnras/stz521. arXiv: 1804.07341 [astro-ph.HE].

Bogovalov, S. V., D. Khangulyan, A. Koldoba, G. V. Ustyugova, and F. Aharonian. 2019. Modelling the interaction between relativistic and non-relativistic winds in binary pulsar systems: strong magnetization of the pulsar wind. *MNRAS* 490, no. 3 (December): 3601–3607. https://doi.org/10.1093/mnras/stz2815. arXiv: 1911.07441 [astro-ph.HE].

Bogovalov, S. V., D. Khangulyan, A. V. Koldoba, G. V. Ustyugova, and F. A. Aharonian. 2012. Modelling the interaction between relativistic and non-relativistic winds in the binary system PSR B1259-63/SS2883- II. Impact of the magnetization and anisotropy of the pulsar wind. *MNRAS* 419, no. 4 (February): 3426–3432. https://doi.org/10.1111/j.1365-2966.2011.19983.x. arXiv: 1107.4831 [astro-ph.HE].

Bogovalov, S. V., D. V. Khangulyan, A. V. Koldoba, G. V. Ustyugova, and F. A. Aharonian. 2008. Modelling interaction of relativistic and non-relativistic winds in binary system PSR B1259-63/SS2883 - I. Hydrodynamical limit. *MNRAS* 387, no. 1 (June): 63–72. https://doi.org/10.1111/j.1365-2966.2008.13226.x. arXiv: 0710.1961 [astro-ph].

Bosch-Ramon, V., and M. V. Barkov. 2011. Large-scale flow dynamics and radiation in pulsar γ-ray binaries. *A&A* 535 (November): A20. https://doi.org/10.1051/0004-6361/201117235. arXiv: 1105.6236 [astro-ph.HE].

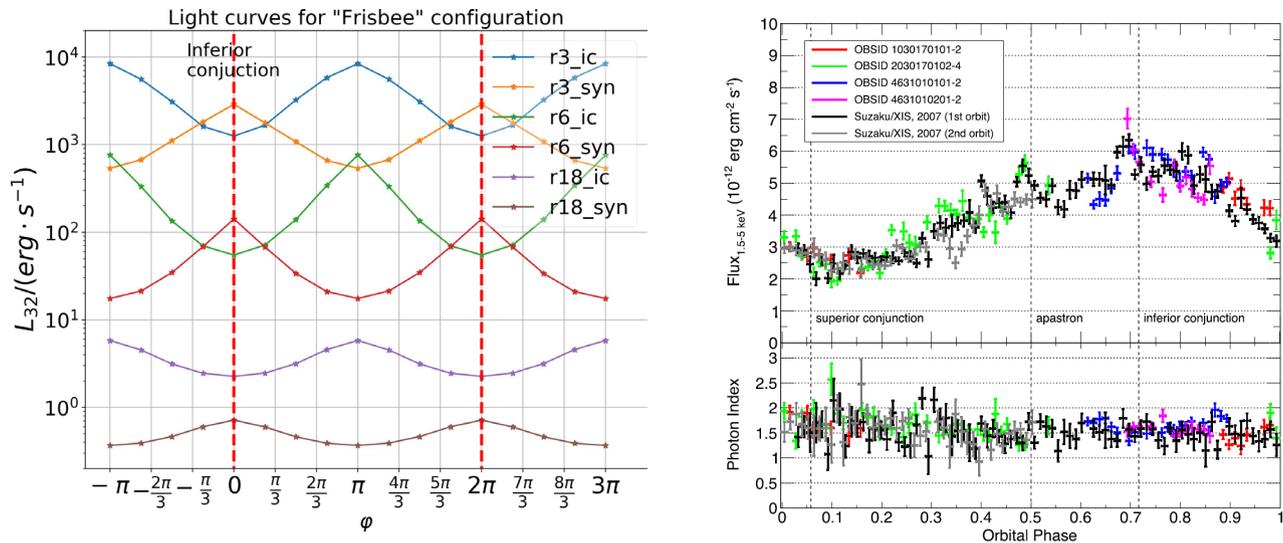

**Figure 20.** Light curves of $\theta = \pi/6$ for "Frisbee" configuration on the left and LS5039 X-ray observed light curve on the right (we take Fig. 1 from Yoneda et al. 2023). The left panel has six light curves, which has changed the $\eta$-parameter ($\eta = 1/4$ for r3, $\eta = 1/25$ for r6, and $\eta = 1/289$ for r18) and types of radiation (SYN or IC). The orbital angle $\varphi$ changes from $-\pi$ to $3\pi$. The $\varphi = 0$ corresponds to the neutron star inferior conjunction.

Bosch-Ramon, V., M. V. Barkov, D. Khangulyan, and M. Perucho. 2012. Simulations of stellar/pulsar-wind interaction along one full orbit. *A&A* 544 (August): A59. https://doi.org/10.1051/0004-6361/201219251. arXiv: 1203.5528 [astro-ph.HE].

Bosch-Ramon, V., M. V. Barkov, and M. Perucho. 2015. Orbital evolution of colliding star and pulsar winds in 2D and 3D: effects of dimensionality, EoS, resolution, and grid size. *A&A* 577 (May): A89. https://doi.org/10.1051/0004-6361/201425228. arXiv: 1411.7892 [astro-ph.HE].

Bosch-Ramon, Valenti, Maxim V. Barkov, Andrea Mignone, and Pol Bordas. 2017. HESS J0632+057: hydrodynamics and non-thermal emission. *MNRAS* 471, no. 1 (October): L150–L154. https://doi.org/10.1093/mnrasl/slx124. arXiv: 1708.00066 [astro-ph.HE].

Bosch-Ramon, Valenti, and Dmitry Khangulyan. 2009. Understanding the Very-High Emission from Microquasars. *International Journal of Modern Physics D* 18, no. 3 (January): 347–387. https://doi.org/10.1142/S0218271809014601. arXiv: 0805.4123 [astro-ph].

Casares, J, M Ribo, I Ribas, JM Paredes, Javier Marti, and Artemio Herrero. 2005. A possible black hole in the γ-ray microquasar ls 5039. *Monthly Notices of the Royal Astronomical Society* 364 (3): 899–908.

Casares, J., M. Ribo, I. Ribas, J. M. Paredes, J. Marti, and A. Herrero. 2005. A possible black hole in the γ-ray microquasar LS 5039. *MNRAS* 364, no. 3 (December): 899–908. https://doi.org/10.1111/j.1365-2966.2005.09617.x. arXiv: astro-ph/0507549 [astro-ph].

Collmar, W., and S. Zhang. 2014. LS 5039 - the counterpart of the unidentified MeV source GRO J1823-12. *A&A* 565 (May): A38. https://doi.org/10.1051/0004-6361/201323193. arXiv: 1402.2525 [astro-ph.HE].

Dubus, G. 2006. Gamma-ray binaries: pulsars in disguise? *A&A* 456, no. 3 (September): 801–817. https://doi.org/10.1051/0004-6361:20054779. arXiv: astro-ph/0605287 [astro-ph].

Dubus, G., A. Lamberts, and S. Fromang. 2015. Modelling the high-energy emission from gamma-ray binaries using numerical relativistic hydrodynamics. *A&A* 581 (September): A27. https://doi.org/10.1051/0004-6361/201425394. arXiv: 1505.01026 [astro-ph.HE].

Dubus, Guillaume. 2013. Gamma-ray binaries and related systems. *A&A Rev.* 21 (August): 64. https://doi.org/10.1007/s00159-013-0064-5. arXiv: 1307.7083 [astro-ph.HE].

Hadasch, D., D. F. Torres, T. Tanaka, R. H. D. Corbet, A. B. Hill, R. Dubois, G. Dubus, et al. 2012. Long-term Monitoring of the High-energy γ-Ray Emission from LS I +61°303 and LS 5039. *ApJ* 749, no. 1 (April): 54. https://doi.org/10.1088/0004-637X/749/1/54. arXiv: 1202.1866 [astro-ph.HE].

Harten, Ami. 1983. High resolution schemes for hyperbolic conservation laws. *Journal of Computational Physics* 49 (3): 357–393. ISSN: 0021-9991. https://doi.org/10.1016/0021-9991(83)90136-5. http://www.sciencedirect.com/science/article/pii/0021999183901365.

Huber, D., R. Kissmann, and O. Reimer. 2021. Relativistic fluid modelling of gamma-ray binaries. II. Application to LS 5039. *A&A* 649 (May): A71. https://doi.org/10.1051/0004-6361/202039278. arXiv: 2103.00995 [astro-ph.HE].

Khangulyan, D., F. Aharonian, and V. Bosch-Ramon. 2008. On the formation of TeV radiation in LS 5039. *MNRAS* 383, no. 2 (January): 467–478. https://doi.org/10.1111/j.1365-2966.2007.12572.x. arXiv: 0707.1689 [astro-ph].

Khangulyan, D., F. A. Aharonian, and S. R. Kelner. 2014. Simple Analytical Approximations for Treatment of Inverse Compton Scattering of Relativistic Electrons in the Blackbody Radiation Field. *ApJ* 783, no. 2 (March): 100. https://doi.org/10.1088/0004-637X/783/2/100. arXiv: 1310.7971 [astro-ph.HE].

Khangulyan, D., Maxim V. Barkov, and S. B. Popov. 2022. Fast Radio Bursts by High-frequency Synchrotron Maser Emission Generated at the Reverse Shock of a Powerful Magnetar Flare. *ApJ* 927, no. 1 (March): 2. https://doi.org/10.3847/1538-4357/ac4bdf. arXiv: 2106.09858 [astro-ph.HE].

Khangulyan, D., S. Hnatic, F. Aharonian, and S. Bogovalov. 2007. TeV light curve of PSR B1259–63/SS2883. *Monthly Notices of the Royal Astronomical Society* 380, no. 1 (August): 320–330. ISSN: 0035-8711. https://doi.org/10.1111/j.1365-2966.2007.12075.x. eprint: https://academic.oup.com/mnras/article-pdf/380/1/320/4151107/mnras0380-0320.pdf. https://doi.org/10.1111/j.1365-2966.2007.12075.x.

Komissarov, S. S. 2013. Magnetic dissipation in the Crab nebula. *MNRAS* 428 (January): 2459–2466. https://doi.org/10.1093/mnras/sts214. arXiv: 1207.3192 [astro-ph.HE].

Kong, S. W., K. S. Cheng, and Y. F. Huang. 2012. MODELING THE MULTIWAVELENGTH LIGHT CURVES OF PSR b1259-63/LS 2883. II. THE EFFECTS OF ANISOTROPIC PULSAR WIND AND DOPPLER BOOSTING. *The Astrophysical Journal* 753, no. 2 (June): 127. https://doi.org/10.1088/0004-637x/753/2/127. https://doi.org/10.1088/0004-637x/753/2/127.

Lamberts, A., S. Fromang, G. Dubus, and R. Teyssier. 2013. Simulating gamma-ray binaries with a relativistic extension of RAMSES. *A&A* 560 (December): A79. https://doi.org/10.1051/0004-6361/201322266. arXiv: 1309.7629 [astro-ph.IM].

Lopez-Miralles, J., M. Perucho, J. M. Marti, S. Migliari, and V. Bosch-Ramon. 2022. 3D RMHD simulations of jet-wind interactions in high-mass X-ray binaries. A&A 661 (May): A117. https://doi.org/10.1051/0004-6361/202142968. arXiv: 2202.11119 [astro-ph.HE].

Lyubarsky, Y. E. 2003. The termination shock in a striped pulsar wind. *MNRAS* 345 (October): 153–160. https://doi.org/10.1046/j.1365-8711.2003.06927.x. eprint: astro-ph/0306435.

Michel, F. C. 1973. Rotating Magnetosphere: a Simple Relativistic Model. *ApJ* 180 (February): 207–226.

Mignone, A., G. Bodo, S. Massaglia, T. Matsakos, O. Tesileanu, C. Zanni, and A. Ferrari. 2007. PLUTO: A Numerical Code for Computational Astrophysics. ApJS 170, no. 1 (May): 228–242. https://doi.org/10.1086/513316. arXiv: astro-ph/0701854 [astro-ph].

Molina, E., and V. Bosch-Ramon. 2020. A dynamical and radiation semi-analytical model of pulsar-star colliding winds along the orbit: Application to LS 5039. A&A 641 (September): A84. https://doi.org/10.1051/0004-6361/202038417. arXiv: 2007.00543 [astro-ph.HE].

Olmi, B., and N. Bucciantini. 2019. Full-3D relativistic MHD simulations of bow shock pulsar wind nebulae: emission and polarization. MNRAS 488, no. 4 (October): 5690–5701. https://doi.org/10.1093/mnras/stz2089.arXiv: 1907.12356 [astro-ph.HE].

Porth, O., S. S. Komissarov, and R. Keppens. 2014. Three-dimensional magnetohydrodynamic simulations of the Crab nebula. MNRAS 438 (February): 278–306. https://doi.org/10.1093/mnras/stt2176. arXiv: 1310.2531 [astro-ph.HE].

Sierpowska, A., and W. Bednarek. 2005. γ-rays from cascades in close massive binaries containing energetic pulsars. Monthly Notices of the Royal Astronomical Society 356, no. 2 (January): 711–726. ISSN: 0035-8711. https://doi.org/10.1111/j.1365-2966.2004.08490.x. eprint: https://academic.oup.com/mnras/article-pdf/356/2/711/3979006/356-2-711.pdf. https://doi.org/10.1111/j.1365-2966.2004.08490.x.

Tavani, Marco, and Jonathan Arons. 1997. Theory of High-Energy Emission from the Pulsar/Be Star System PSR 1259-63. I. Radiation Mechanisms and Interaction Geometry. ApJ 477, no. 1 (March): 439–464. https://doi.org/10.1086/303676. arXiv: astro-ph/9609086 [astro-ph].

Yoneda, Hiroki, Valenti Bosch-Ramon, Teruaki Enoto, Dmitry Khangulyan, Paul S. Ray, Tod Strohmayer, Toru Tamagawa, and Zorawar Wadiasingh. 2023. Unveiling Properties of the Nonthermal X-Ray Production in the Gamma-Ray Binary LS 5039 Using the Long-term Pattern of Its Fast X-Ray Variability. ApJ 948, no. 2 (May): 77. https://doi.org/10.3847/1538-4357/acc175. arXiv: 2303.12587 [astro-ph.HE].

Zabalza, V., V. Bosch-Ramon, F. Aharonian, and D. Khangulyan. 2013. Unraveling the high-energy emission components of gamma-ray binaries. A&A 551 (March): A17. https://doi.org/10.1051/0004-6361/201220589. arXiv: 1212.3222 [astro-ph.HE].

**Appendix 1. Pulsar wind structure**

It was assumed that the alternating components of the magnetic field in its striped zone were completely dissipated along its way. Possibly the dissipation has place at the termination shock, but it is not changing the dynamics of the flow see Lyubarsky 2003. For the total energy flux density of the wind, we adopt the monopole model (Michel 1973)

$$f_{tot}(r, \theta_p) = L_0 \left(\frac{1}{r}\right)^2 (\sin^2 \theta_p + g), \qquad (6)$$

to avoid vanishing energy flux at the poles, we add the parameter g = 0.03. This energy is distributed between the magnetic fm component

$$f_m(r, \theta_p) = \frac{\sigma(\theta_p) f_{tot}(r, \theta_p)}{1 + \sigma(\theta_p)}, \qquad (7)$$

and kinetic $f_k$ one

$$f_k(r, \theta_p) = \frac{f_{tot}(r, \theta_p)}{1 + \sigma(\theta_p)}, \qquad (8)$$

where $\sigma(\theta_p)$ is wind magnetization, which depends on latitude, the angle $\theta_p$) is counted from the pulsar spin axis.

For numerical stability magnetization is vanishing at the pole region as

$$\sigma_0(\theta_p) = \sigma_0 \min\left(1, \left(\frac{\theta_p}{\theta_0}\right)^2\right), \qquad (9)$$

here $\theta_0$ is another small parameter which is equal to 0.2. The magnetic stripes dissipation changes the wind magnetization at the equatorial zone as

$$\sigma(\theta_p) = \frac{\sigma_0(\theta_p) \chi_\alpha(\theta_p)}{1 + \sigma_0(\theta_p)\left(1 - \chi_\alpha(\theta_p)\right)}, \qquad (10)$$

where

$$\chi_\alpha(\theta_p) = \begin{cases} (2\phi_\alpha(\theta_p)/\pi - 1)^2, & \frac{\pi}{2} - \alpha < \theta_p < \frac{\pi}{2} + \alpha \\ 1, & otherwise \end{cases}, \qquad (11)$$

and $\phi_\alpha(\theta_p) = arccos(-cot(\theta_p)cot(\alpha))$. The $\alpha$ is the angle between the magnetic axis and the pulsar rotation axis (see for more details in Komissarov 2013).

**Appendix 2. Radiation processes and Doppler-boosting**

The main non-thermal processes dominating the radiation of the interaction of winds are the synchrotron (SYN) and the inverse-Compton (IC). To quantify these processes, cooling times are calculated.

For electrons interacting with stellar photons due to IC process, a characteristic cooling time is used (Bosch-Ramon and Khangulyan 2009; Khangulyan, Aharonian, and Kelner 2014; Barkov and Bosch-Ramon 2018), expressed as:

$$t_{IC} = \gamma/\dot{\gamma}_{IC}, \quad (12)$$

$$\dot{\gamma}_{IC} = 5.5 \times 10^{17} \, T_{mcc}^3 \gamma \log_{10}(1 + 0.55 \, \gamma \, Tmcc) \quad (13)$$

$$\times \frac{1 + \dfrac{1.4 \, \gamma T_{mcc}}{1 + 12\gamma^2 T_{mcc}^2}}{1 + 25\gamma T_{mcc}} \left(\frac{R_*}{r}\right)^2,$$

where $\gamma$ is the electron Lorentz factor, $T_{mcc} = kT_*/m_e c^2$, the stellar temperature in electron rest energy units, $R_*$ the stellar radius, and r the distance to the star from the radiating region.

For SYN radiation, the time characteristic has the form:

$$t_{SYN} \approx \frac{6 \times 10^2}{B^2 E_{SYN}}, \, s, \quad (14)$$

where B is the magnetic field in G, and

$$E_{SYN} \approx 0.2 \left(\frac{\epsilon_{\gamma,keV}}{B[G]}\right)^{1/2}, \, erg, \quad (15)$$

The characteristic value of the emissivity of one calculation cell can be estimated as $u_{cell}/t_{rad}$, where $u_{cell} = 3P$ and subscript "rad" means SYN or IC. But to calculate the luminosity of the entire system, it is necessary to sum up the resulting value over the entire volume $L_{rad} \sim \int_V (u_{cell}/t_{rad}) dV$.

The obtained values refer to the radiation in its own frame of reference. To find quantitative values of the emissivity on the part of the observer (laboratory reference frame), it is necessary to take into account the change in the frequency of the emitted photon $\nu' = \nu/\delta$ and the Doppler boosting, where the Doppler factor is $\delta = 1/\Gamma(1-\beta \cos \theta_{obs})$, $\beta = v/c$ and $\theta_{obs}$ is the angle between the directions of the observer and the motion of commoving frame of plasma. The detailed procedure of Doppler boosting calculation was taken from (Barkov, Lyutikov, and Khangulyan 2019).